\documentclass[authoryear,3p]{elsarticle}

\journal{Alexandria Engineering Journal}

\usepackage{graphicx}

\usepackage{amssymb,amsmath}
\usepackage{enumitem}
\usepackage{mathtools}
\usepackage{multirow}
\usepackage{comment}
\usepackage[graphicx]{realboxes}
\usepackage{url}
\usepackage[hyperindex,pdftex]{hyperref}

\usepackage{xcolor}

\usepackage{comment}
\usepackage{algorithm,algorithmic}
\usepackage{placeins}

\newdefinition{definition}{Definition}
\newdefinition{assumption}{Assumption}
\newdefinition{remark}{Remark}

\begin{document}

\begin{frontmatter}




  \title{The COVID-19 (SARS-CoV-2) Uncertainty Tripod in Brazil: Assessments on model-based predictions with large under-reporting}

\author[1,2]{Saulo B. Bastos}

\author[3]{Marcelo M. Morato}
\ead{marcelomnzm@gmail.com}

\author[1,2,4]{Daniel O. Cajueiro}
\ead{danielcajueiro@unb.br}

\author[3]{Julio E. Normey-Rico}
\ead{julio.normey@ufsc.br}

\address[1]{Departamento de Economia, FACE, Universidade de Bras\'{i}lia (UnB), \\ Campus Universit\'{a}rio Darcy Ribeiro, 70910-900, Bras\'{i}lia, Brazil.}

\address[2]{Machine Learning Laboratory in Finance and Organizations (\emph{LAMFO}), FACE, Universidade de Bras\'{i}lia (UnB), \\ Campus Universit\'{a}rio Darcy Ribeiro, 70910-900, Bras\'{i}lia, Brazil.}

\address[3]{Renewable Energy Research Group (\emph{GPER}), Departamento
  de Automa\c{c}\~ao e Sistemas (\emph{DAS}), \\Universidade Federal de Santa Catarina (UFSC),
  Florian\'opolis, Brazil.}

\address[4]{Nacional Institute of Science and Technology for Complex Systems (INCT-SC).}


\begin{abstract}
The COVID-19 pandemic (SARS-CoV-2 virus) is the defying global health crisis of our time. The absence of mass testing and the relevant presence of asymptomatic individuals causes the available data of the COVID-19 pandemic in Brazil to be largely under-reported regarding the number of infected individuals and deaths. We propose an adapted Susceptible-Infected-Recovered (SIR) model which explicitly incorporates the under-reporting and the response of the population to public policies (such as confinement measures, widespread use of masks, etc) to cast short-term and long-term predictions. Large amounts of uncertainty could provide misleading models and predictions. In this paper, we discuss the role of uncertainty in these prediction, which is illustrated regarding three key aspects. First, assuming that the number of infected individuals is under-reported, we demonstrate an anticipation regarding the peak of infection. Furthermore, while a model with a single class of infected individuals yields forecasts with increased peaks, a model that considers both symptomatic and asymptomatic infected individuals suggests a decrease of the peak of symptomatic. Second, considering that the actual amount of deaths is larger than what is being register, then demonstrate the increase of the mortality rates. Third, when consider generally under-reported data, we demonstrate how the transmission and recovery rate model parameters change qualitatively and quantitatively.  We also investigate the effect of the ``COVID-19 under-reporting tripod'', i.e. the under-reporting in terms of infected individuals, of deaths and the true mortality rate. If two of these factors are known, the remainder can be inferred, as long as proportions are kept constant. The proposed approach allows one to determine the margins of uncertainty by assessments on the observed and true mortality rates.

\end{abstract}


\end{frontmatter}


\section{Introduction}

The COVID-19 pandemic is definitely the global crisis of our time. A Chinese scientist first identified the SARS-CoV-2 virus in humans in Wuhan, in the province of Hubei, China by December 2019. This virus causes severe acute respiratory syndromes which can become potentially fatal. By the end of June, the World Healthy Organization (WHO) estimated that the number confirmed cases was already reaching the order $10,000,000$, with over $490,000$ confirmed deaths.

Much more than presenting drastic effects on health systems, social and economical backlashes are already felt by many countries; this is especially evident in countries with larger social inequalities, such as Brazil. The effects of the virus on populations with poorer access to health systems and sanitation facilities are strikingly harder  \citep{kabir2020covid,san2020covid, khachfe2020epidemiological}. The city of S\~ao Paulo shows a very illustrative example of these differences: the city hall released a technical note by the end of April\footnote{Secretaria Municipal da Sa\'{u}de de S\~{a}o Paulo. \textbf{Boletim Quinzenal Covid-19}. April 30, 2020. Available at: \url{https://www.prefeitura.sp.gov.br/cidade/secretarias/upload/saude/PMSP_SMS_COVID19_Boletim\%20Quinzenal_20200430.pdf}.} stating that the observed mortality rate is $10$ times larger in neighborhoods of the city with worse social conditions and precarious housing.

In this paper, we consider the Brazilian COVID-19 pandemic context, as detailed by \cite{werneck2020covid}. Brazil is currently facing many issues due to the SARS-CoV-2 contagion, such as the advance of the virus to farthest western cities, away from urban areas, where medical care is somehow less present. The country has $26$ federated states, which have been choosing different social distancing measures since mid-March\footnote{Throughout this paper, the Year/Month/Day notation is used.}. Even though a strong public health system is available in Brazil, many states have been exhibiting near-collapsing conditions since May, with over $95\, \%$ of Intense Care Unit (ICU) hospital beds occupied with COVID-19 patients \citep{CFM2020}. Furthermore, we note that the SARS-CoV-2 is currently posing great threat to indigenous communities, such as the Yanomami and Ye'kwana ethnicities\footnote{The Brazilian 
Socioenvironmental Institute (ISA, \textit{Instituto
    Socioambiental}, see
  \url{https://www.socioambiental.org/en}) has released a technical note \cite{ferrante2020protect} which warns for the contagion of
COVID-19 of up to $40 \, \%$ of Yanomami Indigenous Lands,
amid the states of Amazonas and Roraima and a long the border between
Brazil and Venezuela, due to the presence of approximately 20000
illegal mining prospectors. Datasets regarding the
  COVID-19 spread amid indigenous communities are available in \url{https://covid19.socioambiental.org}.}. Clearly, the situation is border-lining.  




The first official death due to the SARS-CoV-2 virus in Brazil was registered in March $17$, while the first case was officially notified in February $26$. Through inferential statistics, \cite{delatorre2020tracking} acknowledge the fact that community transmission has been ongoing in the state of S\~ao Paulo since the beginning of February (over one month before the first official reports). This points to empirical evidence that the true amount of infected individuals, and possibly registered deaths, are actually very under-reported.

 Due to absence of mass testing, Brazil is currently only accounting for COVID-19 patients with moderate to severe symptoms\footnote{As well as those deceased due to the SARS-CoV-2 virus.}. People with mild or no symptoms are not being accounted for, since, in the beginning of the pandemic, the Ministry of Health oriented to stay home and not to seek medical help if one only displayed mild symptoms. Added to this fact, the SARS-CoV-2 virus shows itself as an asymptomatic contagion for a large number of individuals. For these reasons, the scientific community has been warning for a possibly huge margins of underestimated cases in Brazil \citep{silva2020bayesian, rocha2020expected, delatorre2020tracking}. Some studies, such as \cite{CovidNoBrasil}, point out to the presence of over $700 \, \%$ of under-reported cases. Furthermore, the daily reports (``datasets") disclosed by the Brazilian Ministry of Health\footnote{These datasets comprise the number of infected and deceased patients at the given day.}, only give an impression of the virus contagion in the past, since, in average, a person exhibits acute symptoms only $20$ days of the moment of infection. Through statistical procedures, \cite{paixo2020estimation} have recently confirmed the empirical evidences that the margin of under-reported cases is quite large in the majority of Brazilian states. The state of S\~ao Paulo seems to be the one with less uncertainty regarding the number of deaths, because the data also partially incorporates those deceased due to severe acute respiratory syndromes even without  COVID-19 testing.
 
Therefore, it seems evident that to account for uncertainty is essential to formulate adequate public health policies to address the SARS-CoV-2 virus spread. With respect to this context, this paper intends to model the role of uncertainty in this pandemic in such way that decision makers are able to plan more coherent, and adherent to reality, policies. It is worth mentioning that propositions to address the pandemic through recurrent social isolation periods has been recently assessed through optimal control in \citep{morato2020optimal}.
 
 Table \ref{tab:estimated_underreporting} summarizes the estimates available in the literature regarding the under-reporting levels and true mortality rates for Brazil. In this Table, we evidence the percentage increase of under-reports w.r.t. to reported deaths and cases, as evaluated by prior studies. Most of these works show that, in average, the number of infected individuals could be $3$ to $14$ times higher. Some studies from other countries point out that this number could reach up to $30$\footnote{Naomi Martin. \textbf{Mass. official coronavirus count is 218, but experts say true number could be as high as 6,500}. The Boston Globe. March 17, 2020. Available at: \url{https://www.bostonglobe.com/2020/03/17/metro/mass-official-coronavirus-count-is-197-experts-say-true-number-could-now-be-high-6000/}.}.
 According to technical news-pieces disclosed by \cite{uol2020mortes} and \cite{globo2020mortes}, the amount of under-reports in terms of deaths due to the SARS-CoV-2 virus ranges from 17\% to 122\%. \cite{bib2020estimativas} and \cite{bib2020analisesubnotificacao} estimate the real mortality rate estimated for Brazil to be approximately 1.08\% to 1.11\%, but estimates from the research group \cite{epicovid4RS_3,epicovid4RS_4} from random samples of the Brazilian population point out that this mortality could be ranging from 0.42\% to 0.97\%. Studies from other countries, such as \cite{Khan2020.05.17.20104554} and \cite{baud2020real} show that this true mortality rate varies roughly from $1$ to $5\%$, but \citep{howDeadlyIsCoronavirus} states that recent research are converging to the estimation of the true mortality rate to be between 0.5 to 1\%. \citep{Perez-Saez2020.06.10.20127423} find the true mortality rate to be approximately 0.64\% for Geneva, Switzerland.

\begin{table}
\hspace*{-1.5cm}
\centering
\begin{tabular}{p{3.5cm}cc@{\extracolsep{5pt}}ccc}
\hline \hline
~ & \multicolumn{2}{c}{\textbf{Infected}} & \multicolumn{2}{c}{\textbf{Deaths}} & \textbf{True} \\ \cline{2-3}\cline{4-5}
 \textbf{Source} & \textbf{Reported} & \textbf{$\boldsymbol{\times}$ more} & \textbf{Reported} & \textbf{$\boldsymbol{\times}$ more} & \textbf{mortality rate} \\
\hline
\citep{bib2020estimativas,bib2020analisesubnotificacao} & 6.55\% & 14.2 & 100\% & 0 & 1.08\% to 1.11\% \\
\citep{nois2020} & 7.8\% to 8.1\% & 11.8 to 12.3 & 100\% & 0 & 1.3\% \\
\citep{observatorio2020mortes} & ~ & ~ & 26.9\% to 37.5\% & 1.67 to 2.72 & ~ \\
\citep{globo2020mortes} & ~ & ~ & 85.5\% & 0.17 & ~ \\
\citep{uol2020mortes} & ~ & ~ & 67.5\% & 0.48 & ~ \\
\citep{epicovid4RS_3} & 10\% (5.9\% to 20\%) & 9 (4 to 16) & ~ & ~ & 0.42\% (0.23\% to 0.87\%) \\
\citep{epicovid4RS_4} & 25\% (13.9\% to 41.7\%) & 3 (1.4 to 6.2) & ~ & ~ & 0.97\% (0.84\% to 1.12\%) \\
\citep{CovidNoBrasil} & 12.5\% & 7 & ~ & ~ & ~ \\
\citep{covidsubnot} & 7.4\% & 12.5 & ~ & ~ & ~ \\ \hline
\hline
\end{tabular}
 \caption{Estimates of COVID-19 sub-report levels in Brazil and true mortality rates.}
\label{tab:estimated_underreporting}
\end{table}

Bearing in mind the previous discussion, the main motivation of this paper is to present an adapted Susceptible-Infected-Recovered (SIR) model which inherently takes into account these uncertainty levels, considering the Brazilian COVID-19 context. Furthermore, the motivation is also to assess the role of uncertainty in the predictions cast with such models. We denote uncertainty as the amount of sub-notification with respect to infected and deceased individuals. Our approach comprises the following ingredients:
\begin{itemize}
    \item Firstly, we propose a new modelling scheme that incorporates a dynamic decaying parameter for the viral transmission rate. The dynamic decaying parameter for the transmission rate, adapted from \cite{morato2020optimal}, considers that the government applies contagion mitigation measures (such as social isolation and incentives to use masks, which we refer as ``pandemic policies'' henceforth), which decreases the contagion spread dynamics. 
    
    \item Secondly, we develop an uncertainty measure with respect to infected and deceased individuals. The uncertainty is embedded to the optimization procedures used to determine the epidemiological parameters, in order to correct underestimates of infected and deceased individuals.
    
    \item Thirdly, we use this adapted SIR models to make predictions for the Brazilian scenario regarding several different uncertainty sets, with short and long term forecast spans. By this, we are able to illustrate the role of under-reporting in the model response curves.
    Specifically, we study its effect upon peak of infections (in terms of amplitude and time shift), upon the total number of deaths, observed mortality rate, and model epidemiological parameters.

    \item Finally, considering the uncertainty tripod, i.e. the strong link between under-reporting of infected and deaths and the true mortality rate, we extrapolate and suggest an alignment of the observed and the true mortality rate to infer on the level of uncertainty present on the measurements. 
\end{itemize}


We note that this paper relates to a previous paper by the Authors \citep{bastos2020modeling}, wherein  SIR-like models present short and long term outlooks for Brazil. The previous work covered only the data from an early stage of the contagion, until March 30, 2020. Furthermore, the Authors used parametric variations of the parameters to introduce the uncertainty in the identified model, differing from the approach proposed herein.

\begin{figure}
\begin{center}
\includegraphics[width=\linewidth]{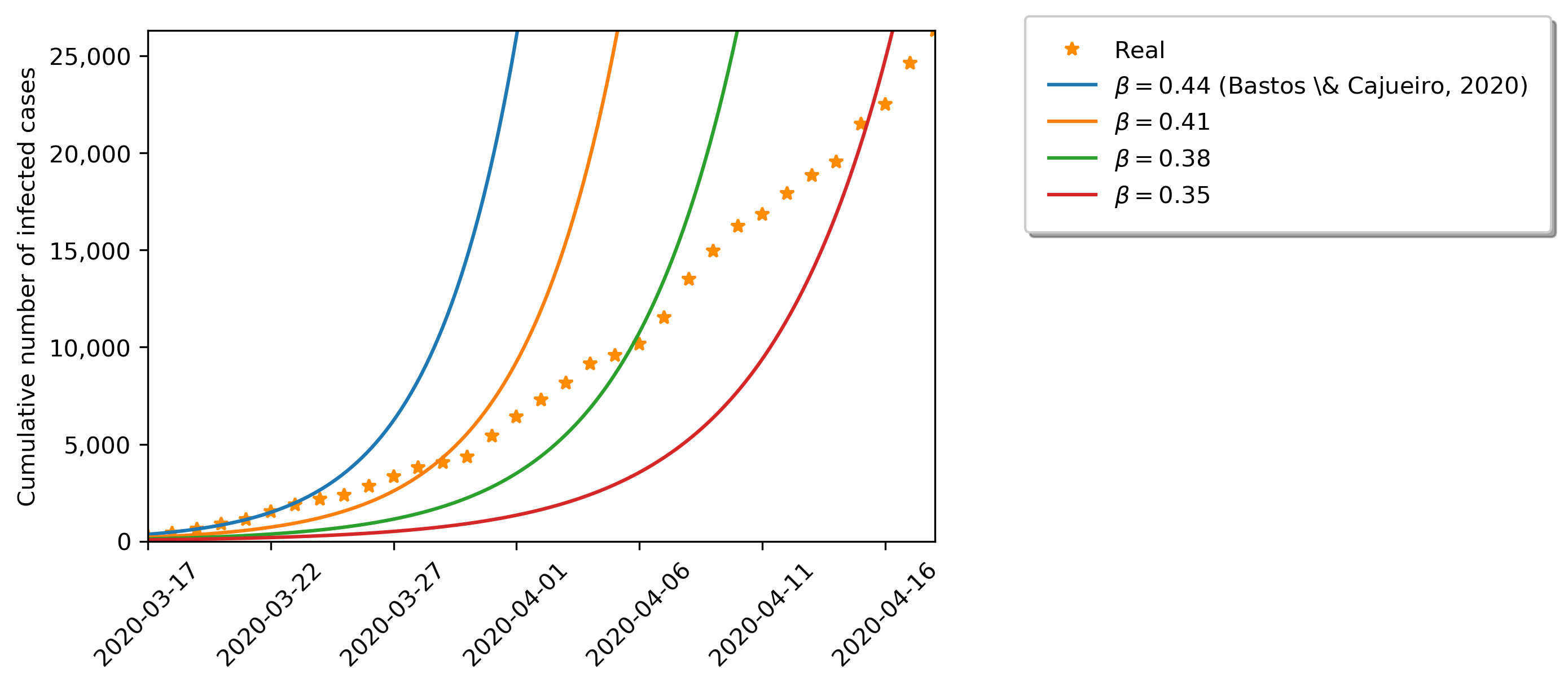}
\caption{\label{fig:cum_infected_does_not_work}Cumulative number of infected.}
\end{center}
\end{figure}

Figure \ref{fig:cum_infected_does_not_work} shows the SIR model forecasts back in March 17, with respect to real data, considering variations for the transmission parameter \(\beta\). We note that parametrically changing the transmission parameter is not enough to make the model fit to the real data. Therefore, a time-varying parameter is added to model the effect of pandemic policies, which meddle with the viral transmission rates. Such adapted model is able to account for the population response to governmental enacted policies. In practice, the pandemic policy can be understood as a feedback of endogenous variables, also depending non-observed time-varying factors (people can simply decide to relax quarantine measures, even if the a social isolation policy is still enacted). Through our simulations, we are able to replicate the real data with a fair amount of similarity, as shown in the sequel.



We organize this paper as follows. Section \ref{sec:models} presents the proposed SIR-adapted epidemiological model, which incorporates a new dynamic variable that describes the response of the population to enacted pandemic policies. In this section, we also discuss uncertainty modelling, in terms of the available datasets. Section \ref{SecIdentif} details the parameter estimation procedure, which inherently includes uncertainty. Section \ref{SecResults} shows the results in terms of parameter estimation and forecasts for the COVID-19 pandemic in Brazil. We present a through discussion on the achieved results and concluding remarks in Section \ref{sec:discussion}.
 
\section{SIR Epidemiological Models}
\label{sec:models}

Recent literature \citep{peng2020epidemic,kucharski2020early} has demonstrated how the the infection rate and evolution dynamics of the SARS-CoV-2 virus can be adequately described by Susceptible-Infected-Recovered kind models. In this Section, we present the classical SIR model due to \cite{Kermack1927}, the new dynamic variable which models the population's response to isolation policies (enacted by local governments) and discuss some remarks on data uncertainty.

\subsection{Epidemiological model}

The SIR describes the spread of a given disease with respect to a population split into three non-intersecting classes, which stand for:
\begin{itemize}
    \item The total amount of susceptible individuals, that are prone to contract the disease at a given moment of time $t$, denoted through the dynamic variable $S(t)$;
    \item The individuals that are currently infected with the disease (active infections at a given moment of time $t$), denoted through the dynamic variable $I(t)$;
    \item The total amount of recovered individuals, that have already recovered from the disease, from an initial time instant $0$ until the current time $t$, denoted through the dynamic variable $R(t)$
\end{itemize}   
Due to the evolution of the spread of the disease, the size of each of these classes change over time and the total population size \(N\) is the sum of these three classes, as follows:
\begin{equation}
N(t)=S(t)+I(t)+R(t)\label{eq:Nconstant} \, \text{.}
\end{equation}

In the SIR model, the parameter \(\beta\) stands for the average number of contacts that are sufficient for transmission of the virus from one individual, per unit of time \(t\). Therefore,  \(\beta I(t)/N(t)\) determines the average number of contacts that are sufficient for transmission from infected individuals, per unit of time, to one susceptible individual; and \((\beta I(t)/N(t))S(t)\) determines the number of new cases per unit of time due to the amount of \(S(t)\) susceptible individuals (they are ``available for infection''). 

Furthermore, the parameter \(\gamma\) stands for the recovery rate, which is the rate that each infected individual recovers (or dies). This parameter characterizes the amount of individuals that ``leaves'' the infected class, considering a constant probability quota per unit of time. We model the amount of deceased individuals due to a SARS-CoV-2 infection following the lines of \citep{keeling2011}, where $D(t)$ is the dynamic variable which describes the amount of deaths and $\rho$ is the observed mortality rate. 

The ``{\bf SIRD}'' (Susceptible-Infected-Recovered-Dead) model is expressed as follows:
\begin{equation}
    \begin{array}{rcl}
    \displaystyle \frac{dS(t)}{dt} & = & \displaystyle - (1-\psi(t)) \frac{\beta I(t) S(t)}{N(t)}\\[3mm]
    \displaystyle \frac{dI(t)}{dt} & = & \displaystyle (1-\psi(t)) \frac{\beta I(t) S(t)}{N(t)} - \frac{\gamma I(t)}{1 - \rho} \\[3mm]
    \displaystyle \frac{dR(t)}{dt} & = & \displaystyle \gamma I(t) \\[3mm]
    \displaystyle \frac{dD(t)}{dt} & = & \displaystyle \frac{\rho}{1-\rho} \gamma I(t) \\[5mm]
    \end{array}\;\;\;\textrm{\bf [SIRD]}
\label{eq:SIRDead}
\end{equation}

 In this model, $\psi$ represents a transmission rate mitigation factor\footnote{We note, with respect to previous works \citep{morato2020optimal,bastos2020modeling}, that the $\psi$ parameter used in this work represents for $1-\psi$ in the prior. We believe the current representation is easier to grasp.}: for $\psi \, = \, 0$, there is "no control" of the viral spread, while for $\psi \, = \, 1$, the contagion is completely controlled, with no more social interactions (a complete lockdown of all population would represent this scenario). It holds that $N(t) = N_0 - D(t)$, where $N_0$ is the initial population size. Remark that, in SIR kind models, $I(t)$ represents the active infections at a given moment, while $D(t)$ represents the total amount of deaths until this given moment; for this reason, it follows that $\frac{dD(t)}{dt}$ is proportionally dependent to $I(t)$.

Since the SIR model is used herein to describe a short-term pandemic outbreak, we do not consider the effects of demographic variations. Despite recent discussion regarding the possibilities of reinfection \citep{del2020covid}, we assume that a recovered individual does not contract the disease twice in the short period of time of this pandemic.



Since, in the case of the SARS-CoV-2 virus, there is a relevant percentage of the infected individuals that are asymptomatic, we progress by splitting the class of infected individuals into the classes of symptomatic ($I_S)$ and asymptomatic individuals ($I_A)$, as suggested by \citep{Robinson2013,Arino2008,Longini2004}. This yields the ``{\bf SIRASD}'' model\footnote{SIRASD stands for Susceptible-Infected-Recovered-Asymptomatic-Symptomatic-Deaths.}:

\begin{equation}
\begin{array}{rcl}
\displaystyle \frac{dS}{dt} & = & \displaystyle - (1-\psi) (\beta_A I_A(t) + \beta_S I_S(t)) \frac{S}{N(t)} \\[3mm]
\displaystyle \frac{dI_A(t)}{dt} & = & \displaystyle (1-\psi) (1-p) (\beta_A I_A(t) + \beta_S I_S(t)) \frac{S}{N(t)} - \gamma_A I_A(t) \\[3mm]
\displaystyle \frac{dI_S(t)}{dt} & = & \displaystyle (1-\psi) p (\beta_A I_A(t) + \beta_S I_S(t)) \frac{S}{N(t)} - \gamma_S I_S(t) - \frac{\rho_S}{1 - \rho_S}\gamma_S I_S(t) \\[3mm]
\displaystyle \frac{dR_A(t)}{dt} & = & \displaystyle \gamma_A I_A(t) \\[3mm]
\displaystyle \frac{dR_S(t)}{dt} & = & \displaystyle \gamma_S I_S(t) \\[3mm]
\displaystyle \frac{dD(t)}{dt} & = & \displaystyle \frac{\rho_S}{1-\rho_S} \gamma_S I_S(t) \\[3mm]
\end{array}\;\;\;\textrm{\bf [SIRASD]}
\label{eq:SIRASD}
\end{equation}

In this model, $\rho_S$ denotes the actual observed mortality rate for the symptomatic class only, while $\rho$ stands for the total observed mortality rate (for both symptomatic and asymptomatic classes). The parameter $p$ is included to represent the percentage of infected individuals which present symptoms; $(1-p)$ denotes the percentage of those without symptoms. 

Two important issues should be commented: \textit{i}) the mortality for asymptomatic individuals is extremely low; for simplicity it is taken as null in the SIRASD model; and \textit{ii}) the observed (real) mortality rates also includes those that died from the contagion while not being accounted for in the available data sets. This is, if there is a number of sub-reported deaths, this number should also influence the mortality rate parameters.

One can evaluate the  instantaneous values for the morality rates directly as follows:
\begin{equation}
\rho_{S}(t) = \frac{D(t)}{I_{S}(t) + R_{S}(t) + D(t)}
\label{eq:rho_S(t)}
\end{equation}
\begin{equation}
\rho(t) = \frac{D(t)}{I_{A}(t) + R_{A}(t) + I_{S}(t) + R_{S}(t) + D(t)}
\label{eq:rho_t}
\end{equation}

Therefore, when the pandemic is over, it follows that $\displaystyle \rho_{S} = \lim_{t \rightarrow \infty}{\rho_{S}(t)} = D(\infty) / (R_{S}(\infty) + D(\infty))$ and $\displaystyle \rho = \lim_{t \rightarrow \infty}{\rho(t)} = D(\infty) / (R_{A}(\infty) + R_{S}(\infty) + D(\infty))$. Due to this fact, it holds that $\displaystyle \frac{\rho_S}{\rho} = \frac{1}{p}$, i.e. $\rho = p \, \rho_S$.

To take into account the effect of public health policies (that are enacted by local governments to ``control'' and mitigate the effects of the COVID-19 pandemic), such as social isolation, incentives to use of masks, etc., we include the dynamic equation for $\psi (t)$ to the SIRD and SIRASD models:
\begin{equation}
\frac{d\psi(t)}{dt} = \displaystyle \left\{ \begin{array}{rl} \alpha (\psi_{\infty} - \psi(t)) & \text{if under the pandemic policies effect,} \\ 0 & \text{otherwise.} \end{array} \right.
\label{eq:Psi}
\end{equation}

Note that $\psi(t)$ converges to $\psi_{\infty}$ with a settling time of $\frac{3}{\alpha}$\footnote{Considering that $\psi(t) = \psi_{\infty}(1 - e^{-\alpha t})$ and $\displaystyle \frac{\psi_{\infty} - \psi_{\tau}}{\psi_{\infty}} = e^{-\alpha \tau} = \delta \Rightarrow \tau = - \frac{\ln{\delta}}{\alpha}$, if $\delta = 0.05$ then $\displaystyle \tau \approx \frac{3}{\alpha}$.}. It follows that $\psi_\infty$ is a factor which stands for the  maximal observed effect of pandemic policies. Note that for larger values of $\psi_\infty$ (closer to $1$), the spread of the SARS-CoV-2 virus gets slower, with smaller peak of infections and number of deaths. The condition of $\psi \, = \, 1$ represents a total isolation condition, where the amount of contacts is reduced to zero.

In the recent paper by the Authors \citep{morato2020optimal}, these new models were used to conceive model-based optimal control policies, under a Predictive Control optimization formalism. Through the sequel, we refer to these extended models (which embed the effect of public health policies) as  SIRD$+\psi$ and SIRASD$+\psi$ models, respectively.

\subsection{Uncertainty Model}

As discussed in the Introduction, the sub-notification regarding the number of infected and deceased individuals, due to COVID-19, has been pointed out as rather elevated in Brazil, as brought to focus by a number of recent papers \citep{ImperialCollegeNew,THELANCET20201461, silva2020bayesian,rocha2020expected,rodriguez2020covid,morato2020optimal}. These studies discuss that the amount of under-reported cases regarding infected individuals could be up to $30$ times the reported values. Furthermore, the sub-notification in terms of deaths due to the SARS-CoV-2 virus might be over $120$ \% the mortality disclosed by the Ministry of Health.

Therefore, we proceed by modelling these uncertainty margins as follows:
\begin{eqnarray}
I_S^{\text{real}}(t) &=& I_S^{\text{measured}}(t) + I_S^{\text{unknown}}(t)  \, \text{,} \\
D^{\text{real}}(t) &=& D^{\text{measured}}(t) + D^{\text{unknown}}(t) \, \text{,}
\end{eqnarray}
where the super-index "measured" denotes the values as prescribed by the Ministry of Health (data), "unknown" as the sub-notified values and "real" for total value, including uncertainty.

It follows that the amount of uncertainty in any of these variables, generically referred to as $X$, can be described by a concentrated multiplicative parameter, this is:

\begin{equation}
\left.
\begin{array}{rcl}
 X^{\text{measured}}(t) & = & q_X X^{\text{real}}(t) \\[3mm]
 X^{\text{unknown}}(t) & = & (1-q_X) X^{\text{real}}(t) \\
\end{array}
\right\}
\Rightarrow
\frac{X^{\text{measured}}(t)}{X^{\text{unknown}}(t)} = \frac{q_X}{1-q_X} \quad \text{,}
\label{eq:d_measured_unknown_relationship}
\end{equation}
with $X$ representing either $D$ or $I_S$ and $q_X$ denoting the respective uncertainty parameter. Such concentrated parameter $q_j \in [0,1]$ gives a measure for the amount of sub-notification. For instance, if we consider $q_D \, = \, 1$, it means that there is no sub-notification with respect to the disclosed datasets for the number of deaths. For $q_D \, = \, 0.5$, we observe that the actual number of deaths is twice the reported amount.

Using the same notation, through the sequel, we present our results using the following transformation, for understanding simplicity:
\begin{equation}
 \tilde{q}_X = \frac{1}{q_X} - 1 \, \text{,}
 \label{eq:q_function}
\end{equation}
\noindent which holds for $X^{\text{unknown}}(t) \,=\, \tilde{q}_X X^{\text{measured}} (t) \, \Rightarrow \, X^{\text{real}}(t) \, =\,  (1 + \tilde{q}_X) X^{\text{measured}}(t)$. This notation concatenates the following idea: as an example, if we consider $\tilde{q}_I \, = \, 0$, it means that the reported amount of infections is equal to the real amount. Consequently, for $\tilde{q}_I \, = \, 0.25$, for instance, it means that the real amount of infections is $25\%$ bigger than the number reported cases.

With respect to the way we model uncertainty, we must stress that this paper is concerned with the exposure of the simulation/prediction phenomena when using uncertainty-embedded identification applied to SIR-like models in order to forecast the COVID-19 pandemic dynamics in Brazil. The discussion resides in the following key points:
 \begin{enumerate}[label = (\alph*)]
    \item the influence of uncertainty in these forecasts;
    \item the temporal shift of the peak of infections according to the level of uncertainty (which would also impact directly in the enacted pandemic policies); 
    \item the temporal aspect of the mortality rate;
    \item the under-reporting tripod.
\end{enumerate}

Furthermore, we must emphasize that the prediction for the amount of deaths can be quite equivocate according to the level of uncertainty that is considered, since the parameters $q_D$ and $q_I$ directly meddle with the "D" output of the SIRASD+$\psi$ model. It seems reasonable that as time progresses and the pandemic contagion gets "controlled" (i.e. stabilizes), the amount of uncertainty tends to decrease. Anyhow, we must assure that the previewed phenomena will irrevocably take place (the peak of infections and a possible second peak), if the enacted pandemic policies remain the same.

\section{Estimation procedure}
\label{SecIdentif}

The Brazilian Ministry of Health provides (daily) two useful data time-series which are used to evaluate (i.e. identify) the SIR model parameters: \textit{i}) the total number of infected individuals, denoted $$I^{\text{total}}(t) \, =\, I(t) + R(t) + D(t)$$ and \textit{ii}) the total number of deaths, which is $D(t)$. In this paper, we use data from February 25, 2020, to May 31, 2020.

Parameters $\beta$, $\gamma$ and $\rho$ are computed according to the procedure adapted from \cite{bastos2020modeling}, considering that the period for which pandemic policies had not yet been formally implemented (this period comprises the weeks from February 25, 2020 until March 22, 2020). We find that the recovery rate $\gamma$ is found to be roughly constant at $0.150876$, which is consistent with other findings in the literature \citep{Read2020, Nishiura2020}. We maintain this value for $\gamma$ throughout the following identification steps. Note that this parameters is inherent to the disease and time-invariant.




In order to estimate the parameters of the SIRD+$\psi$ and SIRASD+$\psi$ models, considering the remainder of the available data (from March 22, 202 onward), we proceed by performing a Least-Squares identification procedure, with respect to different levels of uncertainty ($\tilde{q}_D$ and $\tilde{q}_I$). These uncertainty levels are directly embedded to the datasets prior to the actual identification. Minimizing the square-error between the integrated variables and their real values is in accordance with  regular identification methodologies \citep{Bard1974,Brauer2019}. Then, we proceed by following an hierarchical procedure, as done in \cite{bastos2020modeling}: firstly, we estimate parameters of the SIRD model, then, assuming that the infected individuals in the SIRD model are the symptomatic individuals in the SIRASD model, we estimate the remaining parameters of the SIRASD model.

Our identification procedure also includes constraints \footnote{If more restrictive bounds are used, we mention them  explicitly.} to the possible parameter values: $\beta, \beta_S, \beta_A \in [1/20, 2]$, $\gamma, \gamma_S, \gamma_A \in [1/14, 1/2]$, $\rho, \rho_S \in [0.001, 0.2]$, $\alpha \in [0, 1]$ and $\psi_{\infty} \in [0.0, 1.0]$. These limits are in accordance with those presented by \cite{werneck2020covid} and \cite{bastos2020modeling}. 


Globally, the identification procedure for the SIRD+$\psi$ model resides in solving the following minimization problem:

\begin{equation}
\min_{\beta, \rho, \alpha, \psi_{\infty}} \, \frac{1}{2} \sum_{t}{
\left( \mathcal{L}_I^2 + \mathcal{L}_D^2 \right) }
 \end{equation}
\begin{align}
\mathcal{L}_I &= \log{\left[1 + \left( \frac{I^{\text{total}}(t) - D(t)/q_D}{q_I}\right) \frac{1}{I_{\text{max}}} \right]} - \log{\left(1 +\frac{ \hat{I}(t)+\hat{R}(t)}{I_{\text{max}}} \right)} \label{eq:loss_I} \\
 \mathcal{L}_D &= \log{\left[1 + \left( \frac{D(t)}{q_D} \right) \frac{1}{D_{\text{max}}} \right]} - \log{\left(1 + \frac{\hat{D}(t)}{D_{\text{max}}} \right)} \label{eq:loss_D}
\end{align}
\noindent where $\hat{I}(t)$, $\hat{R}(t)$ and $\hat{D}(t)$ are estimated values of the infected, recovered and deaths, respectively, and  $\displaystyle I_{\text{max}} = \max_t{ \left( \frac{I^{\text{total}}(t) - D(t)/q_D}{q_I}\right) }$, $\displaystyle D_{\text{max}} = \max_t{ \left( \frac{D(t)}{q_D} \right) }$. We decided to normalize the each series by its maximum value and take the logarithm as an approach to balance different exponential characteristics of the infected and deaths series. The initial conditions for the identification procedures are $D_0 = D(0)/q_D$, $I(0) = I^{\text{cum}}(0)-D(0)$, $R(0) = 0$, $S(0) = N_0 - I(0) - R(0) - D(0)$, where $N_0=210147125$ is the initial population according to Brazilian Institute of Geography and Statistics (IBGE). We estimate $\alpha$ and $\psi_{\infty}$ only in the simulation without any uncertainty, and use these values on the other minimization procedures because we believe that these parameters (a response of the society) should be the same, regardless of the uncertainty level.

Equivalently, the identification procedure for the $SIRASD+\psi$ follows through the minimization problem below:
\begin{equation}
\min_{\beta_A, \gamma_A, \rho} \, \frac{1}{2} \sum_{t}{
\left( \mathcal{L}_{I_S}^2 + \mathcal{L}_D^2 \right) }
 \end{equation}
\begin{equation}
\mathcal{L}_{I_S} = \log{\left[1 + \left( \frac{I^{\text{total}}(t) - D(t)/q_D}{q_I}\right) \frac{1}{I_{\text{max}}} \right]} - \log{\left(1 +\frac{ \hat{I}_S(t)+\hat{R}_S(t)}{I_{\text{max}}} \right)} \label{eq:loss_IS} \end{equation}
\noindent where $\hat{I}_S(t)$ and $\hat{R}_S(t)$ are estimated values of the symptomatic infected and recovered, respectively, and the initial conditions are $D(0) = D(0)/q_D$, $I_{S}(0) = I^{\text{cum}}(0) - D(0)$, $I_{A}(0) = I_{S}(0) (1 - q_I) / q_I$, $R_{S}(0) = R_{A}(0) = 0$, $S(0) = N_0 - D(0) - I_{S}(0) - I_{A}(0) - R_{S}(0) - R_{A}(0)$. For this model, we begin the identification procedure with $\beta_S = \beta$, $\gamma_S = \gamma$, where $\beta$ and $\gamma$ are the values directly taken from the identification procedure regarding the $SIRD+\psi$. Hence, we denote the procedure as hierarchical.


We stress, once again, that a large number of different SIRD+$\psi$ and SIRASD+$\psi$ models are obtained through the identification procedures detailed in the prequel. Each one of these models is identified for different level of uncertainty ($q_D$ and $q_I$).





\section{Results}
\label{SecResults}

In this Section, we present the main results of our paper, which comprise the role that sub-notification uncertainty plays in the model-based predictions of the COVID-19 contagion, harshly affecting the outlooks for its evolution spread in Brazil. To have the available data with such large amounts of uncertainty is a very troublesome issue, since public health policies are currently derived through the available datasets, meaning that these policies may be equivocate, such as reducing social isolation policies before adequate time, and thus lead to possibly unwanted phenomena.

The following results comprise predicted forecasts with the SIRD+$\psi$ and SIRASD+$\psi$ models; they have been obtained using Python software. 

\subsection{SIRD+$\psi$ model}

We begin by depicting the results achieved with the SIRD$+\psi$ model.

\begin{figure}
\begin{center}
\includegraphics[width=120mm]{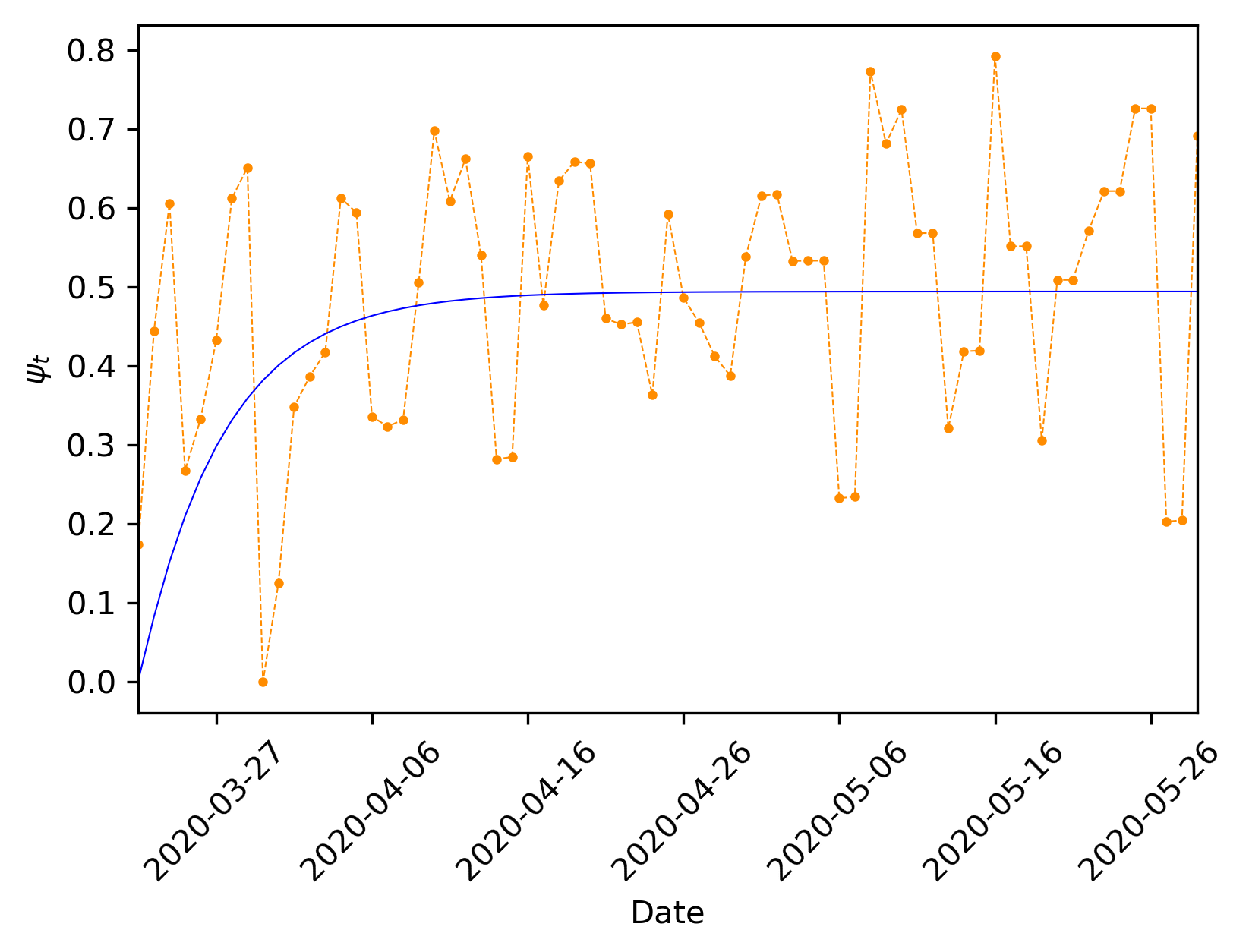}
\caption{\label{fig:estimation_psi}Estimation of $\psi(t)$ for the SIRD$+\psi$ model using the data provided by the Ministry of Health ($\tilde{q}_I = \tilde{q}_D = 0$), which we get $\alpha=0.186353$ and $\psi_{infty}=0.494027$.}
\end{center}
\end{figure}

Figure \ref{fig:estimation_psi} shows the estimation of the $\psi(t)$ variable, which models the population's response to pandemic policies using Eq. (\ref{eq:Psi}), considering no uncertainty ($q_I = q_D = 1$, only for the moment). In this Figure, the orange dots represents the identification results found directly through the $SIRD$ model, as if $\psi$ was a parameter estimated for each sample. The solid blue line represents the time-varying $\psi(t)$ curve. Evidently, we can notice that the estimation of the proposed $\psi (t)$ model absorbs a great deal of error. Consequently, the forecasts derived from such models (SIRD+$\psi$, SIRASD+$\psi$) can vary quite significantly from one day to the next (considering the identification procedure performed daily), while the main trend is followed. Some recent studies \cite{morato2020optimal, Kucharski2020} discuss that the re-estimation (re-identification) should be performed each week, in such way that these daily errors get dynamically absorbed and are thus represented through the mean behaviour. We must re-stress that the following forecasts are extremely sensible to the available datasets (and to the uncertainties) and, thus, to re-perform the identification as time progresses is essential. This is: one cannot use the derived models as if their parameters would not change along time. The forecasts only offer qualitative insights.

We consider several different values for the uncertainty parameters $\breve{q}_I$ and $\breve{q}_D$ in the following simulation runs\footnote{To ease the Reader's interpretation, we transformed $q_I$ and $q_D$ in $\tilde{q}_I$ and $\tilde{q}_D$, respectively, by the use of the transformation in Eq. (\ref{eq:q_function}).}. Anyhow, for exhibition lightness purposes, we will only explicitly explore the ones that found to be more likely scenarios for Brazil (coherent with prior discussion in the literature). Our baseline uncertainty parameters are $\tilde{q}_I = \tilde{q}_D = 0$ (nominal condition), corresponding to the Ministry of Health datasets. We explore $\tilde{q}_I \in \{7, 14, 30\}$ and $\tilde{q}_D \in \{0.25, 0.5, 1\}$, in consonance with previous references. These uncertainty margins relate to the percentage increase regarding infections and deaths, respectively. The gray area presented in the short and long term predictions correspond to uncertainty over the region $\{ (\tilde{q}_I, \tilde{q}_D) | \tilde{q}_I \in [0, 30] ~ \text{and} ~ \tilde{q}_D \in [0, 1] \}$.

\begin{figure}
\centering
   \includegraphics[width=\linewidth]{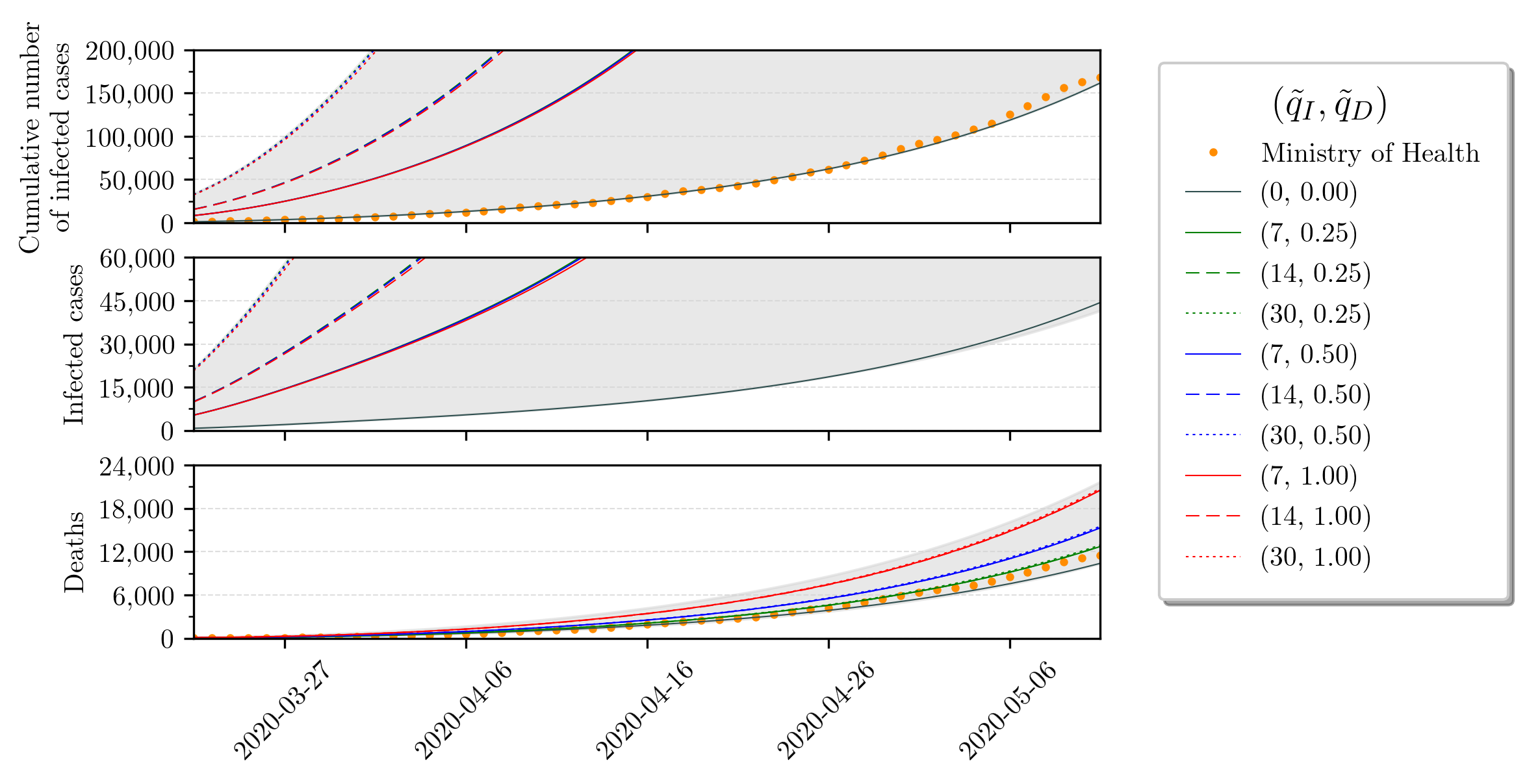}
\caption{Short-term simulation for the SIRD$+\psi$ model using different values of uncertainty.}
\label{fig:simulation_sird_short}
\end{figure}

\begin{figure}
\centering
   \includegraphics[width=\linewidth]{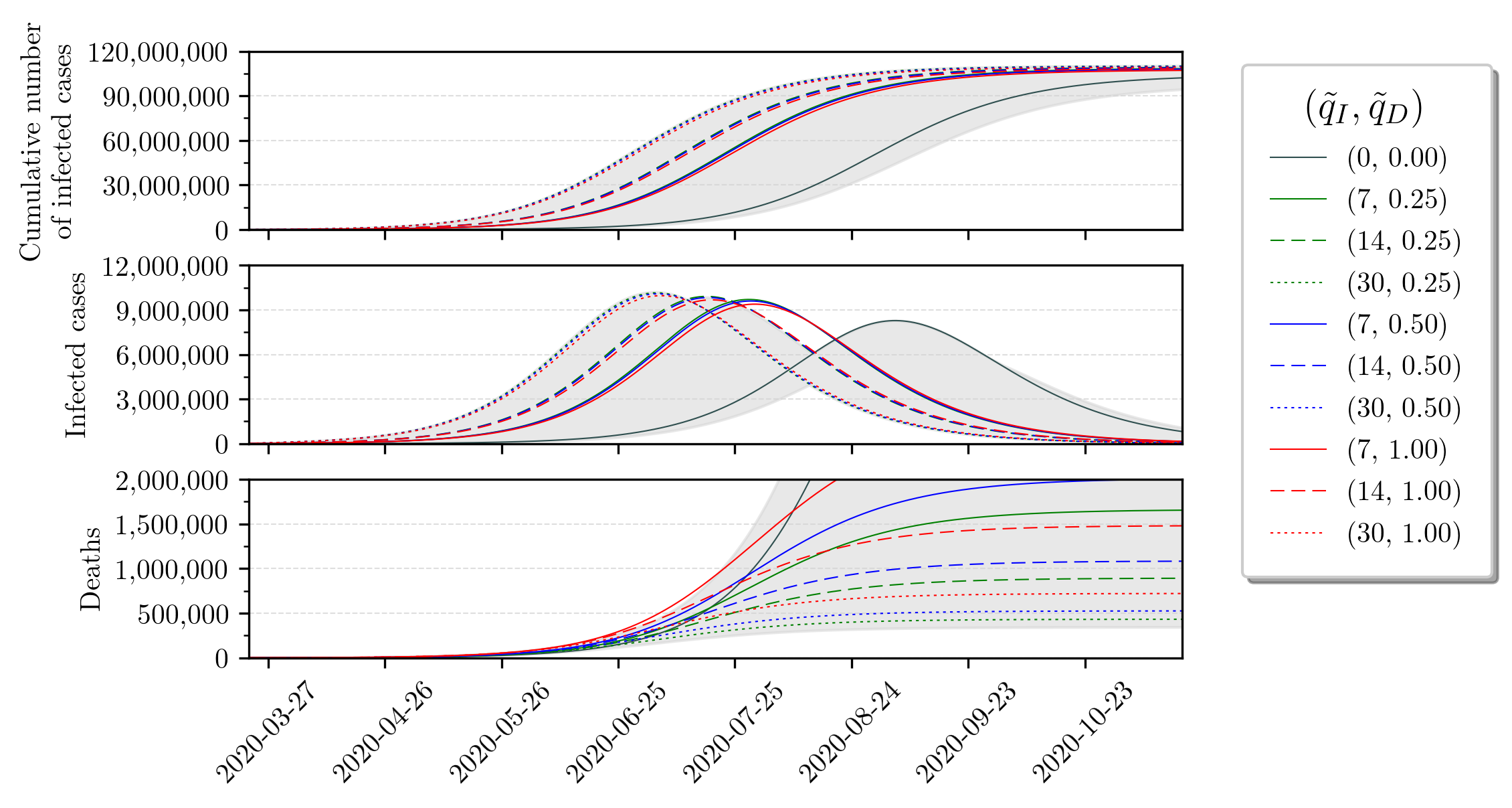}
\caption{Long-term simulation for the SIRD$+\psi$ model using different values of uncertainty.}
\label{fig:simulation_sird_long}
\end{figure}

Based on these uncertainty values, we simulate the SIRD$+\psi$ model along time. We show short and long-term perspectives in Figures \ref{fig:simulation_sird_short} and \ref{fig:simulation_sird_long}. Detailed values that arise in these simulations are concatenated in Table \ref{tab:simulation_info_sird}. With respect to these Figures, we remark the following key points:
\begin{enumerate}
    \item The long-term forecast of the infected cases suggests that greater uncertainty of the number of infected has the effect of anticipating the peak in time and also increase its amplitude. That is, if we have 7, 14 and 30 times more infected, then the peak in September 4 will be anticipated to July 28, 17 and 4, respectively, and also the peak amplitude will increase from 3.9\% to approximately 4.6\%, 4.7\% and 4.8\%, respectively. We note that this peak percentage is given w.r.t. to the total population size.
    \item The peak of infections gets shifted for at least \textbf{one month} for larger uncertainties. This could be quite troublesome, since public health policies concerning ICU beds, for instance, may be accounting for erroneous data and hospitals may be surprised by a larger demand of ICUs than what is expected when disregarding uncertainties. Further analysis are necessary since we cannot separate who are asymptomatic from the ones that are symptomatic and, therefore, would actually demand health care.
    \item The amplitude of the peak of infections shows small variations, despite uncertainties. This suggests that the SARS-CoV-2 will infect the same amount of people. Moreover, the virus does not distinguish between asymptomatic or symptomatic individuals, only aiming to spread its genetic material.
    \item The mortality rate obtained from the Brazilian Ministry of Health datasets does not reflect reality because it differs significantly from the true mortality rate (see Table \ref{tab:estimated_underreporting}). The long-term deaths forecast suggests that the number of infected uncertainty drastically decreases the true mortality rate, while increases in the deaths uncertainty increase the true mortality rate.
\end{enumerate}

\begin{table}
\centering
 \begin{tabular}{cccccc}
\hline
$\tilde{q}_D$ & $\tilde{q}_I$ & $\rho$ & Peak (\%) & Peak Forecast & Number of Deaths \\ \hline
0 & 0 & 8.8569\% & 3.9405\% & September 4, 2020 & 9,226,000 \\
0 & 7 & 1.2086\% & 4.6651\% & July 28, 2020 & 1,316,778 \\
0 & 14 & 0.6447\% & 4.7508\% & July 17, 2020 & 706,295 \\
0 & 30 & 0.3090\% & 4.8573\% & July 4, 2020 & 341,003 \\
0.5 & 7 & 1.8597\% & 4.5681\% & July 29, 2020 & 2,012,722 \\
0.5 & 14 & 0.9958\% & 4.6793\% & July 18, 2020 & 1,085,310 \\
0.5 & 30 & 0.4785\% & 4.7992\% & July 5, 2020 & 525,776 \\
1 & 7 & 2.5450\% & 4.4673\% & July 30, 2020 & 2,735,068 \\
1 & 14 & 1.3683\% & 4.6049\% & July 18, 2020 & 1,483,225 \\
1 & 30 & 0.6594\% & 4.7390\% & July 5, 2020 & 721,160 \\
\hline
\end{tabular}
 \caption{Observed mortality, peak percentage and its respective date of occurrence and number of deaths after 360 days for the SIRD$+\psi$ model for different values of uncertainty.}
\label{tab:simulation_info_sird} 
\end{table}

\begin{figure}
\begin{center}
\includegraphics[width=\linewidth]{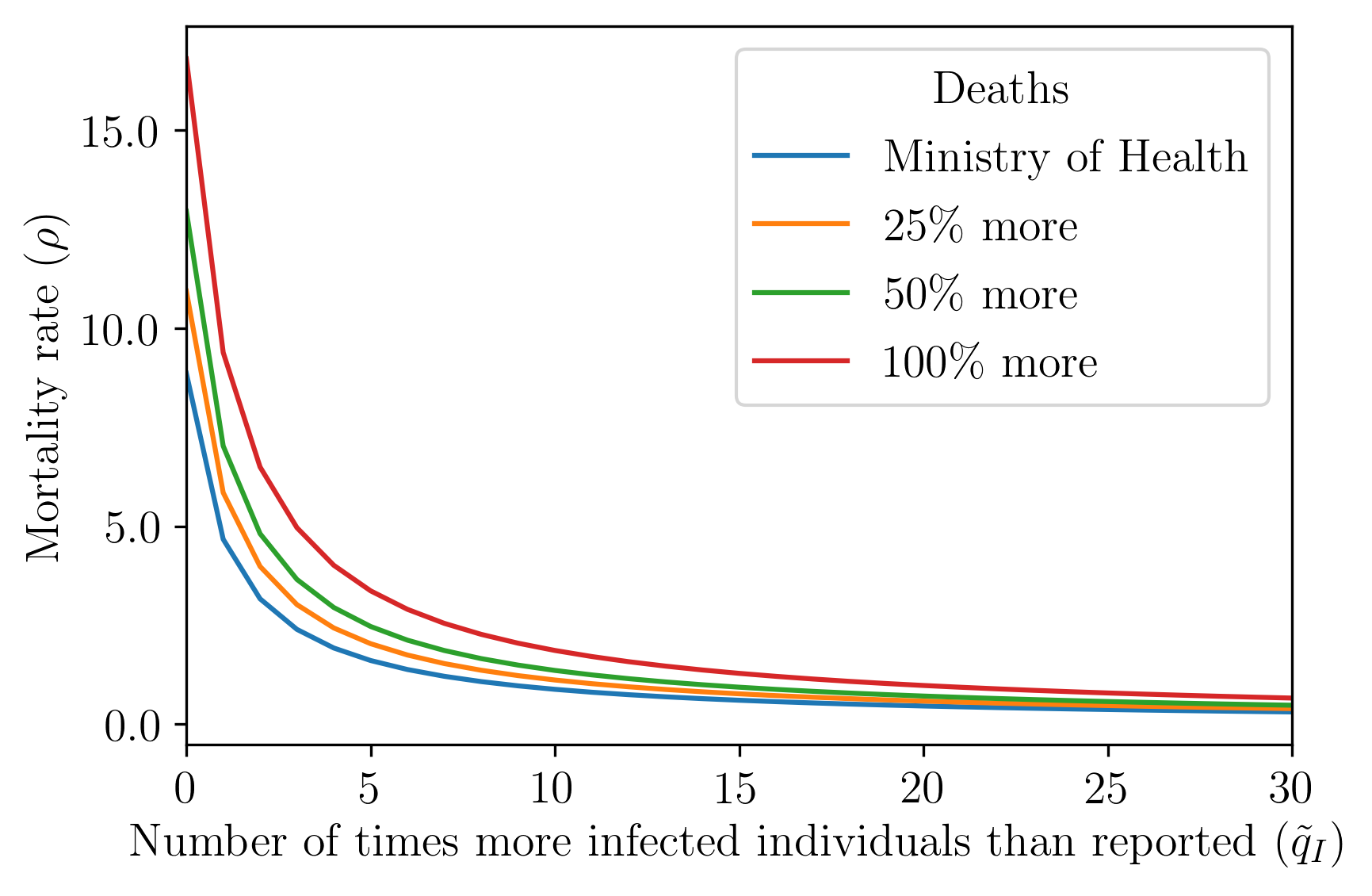}
\caption{\label{fig:mortality_sird}The number of times more infected than reported ($\tilde{q}_I$) vs. the observed mortality rate ($\rho$) for the SIRD$+\psi$ model. The blue, orange and green curves represent simulations using data from the Ministry of Health ($\tilde{q}_D = 0$), 50\% more deaths ($\tilde{q}_D = 0.5$) and 100\% more deaths ($\tilde{q}_D = 1$).}
\end{center}
\end{figure}

We proceed by exploring how the infected uncertainty influences the mortality rate for a given death uncertainty level, as shows Figure \ref{fig:mortality_sird}. In this Figure, we present the estimated values for the mortality rate ($\rho$) from Eq. (\ref{eq:SIRDead}) as a function of the infected uncertainty ($\tilde{q}_I$) for a given (fixed) death uncertainty ($\tilde{q}_D$) amount. The values used in this graph and some estimated epidemiological parameters are in Tables \ref{tab:mortality_sird} and \ref{tab:parameters_sird} in Appendix \ref{sec:appendix_observed_mortality_rates}. We must stress that the observed mortality rate decreases with respect to the increase of infected uncertainty $\tilde{q}_I$; this is shown directly in Figure \ref{fig:mortality_sird}. The reason for this phenomenon resides in the fact that more uncertainty over the infected individuals, while keeping the number of deaths constant, implies that the mortality rate will decrease - see Eq. (\ref{eq:rho_t}).

\begin{figure}
\centering
\includegraphics[width=\linewidth]{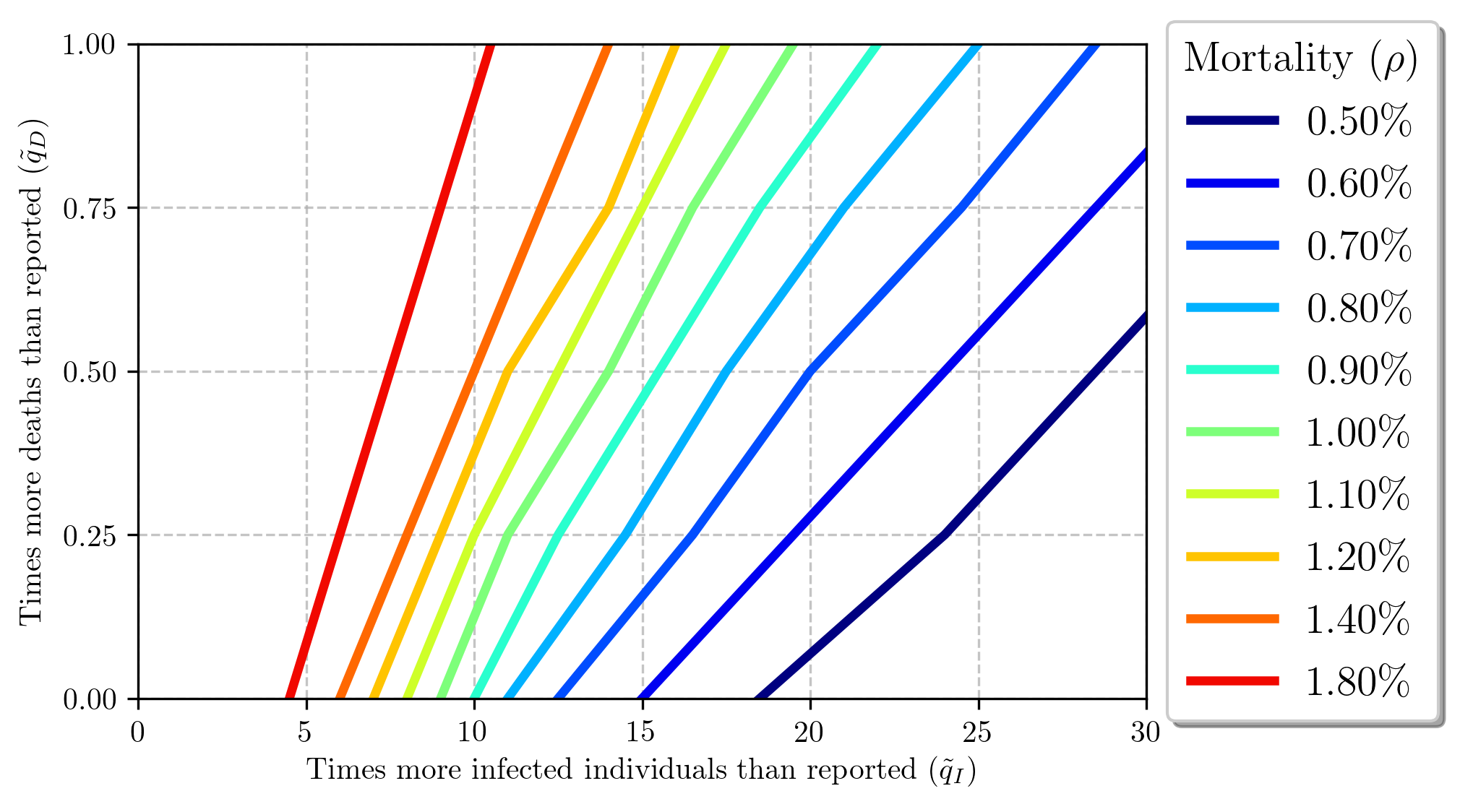}
\caption{Mortality rate paths using the number of times more infected than reported ($\tilde{q}_I$) versus the number of times more deaths than reported ($\tilde{q}_D$) for the SIRD$+\psi$ model.}
\label{fig:sird_mortality_path}
\end{figure}

Figure \ref{fig:sird_mortality_path} provides the uncertainty paths for infected and deaths considering specific mortality rates.

\subsection{SIRASD$+\psi$ model}

\begin{figure}
\centering
   \includegraphics[width=\linewidth]{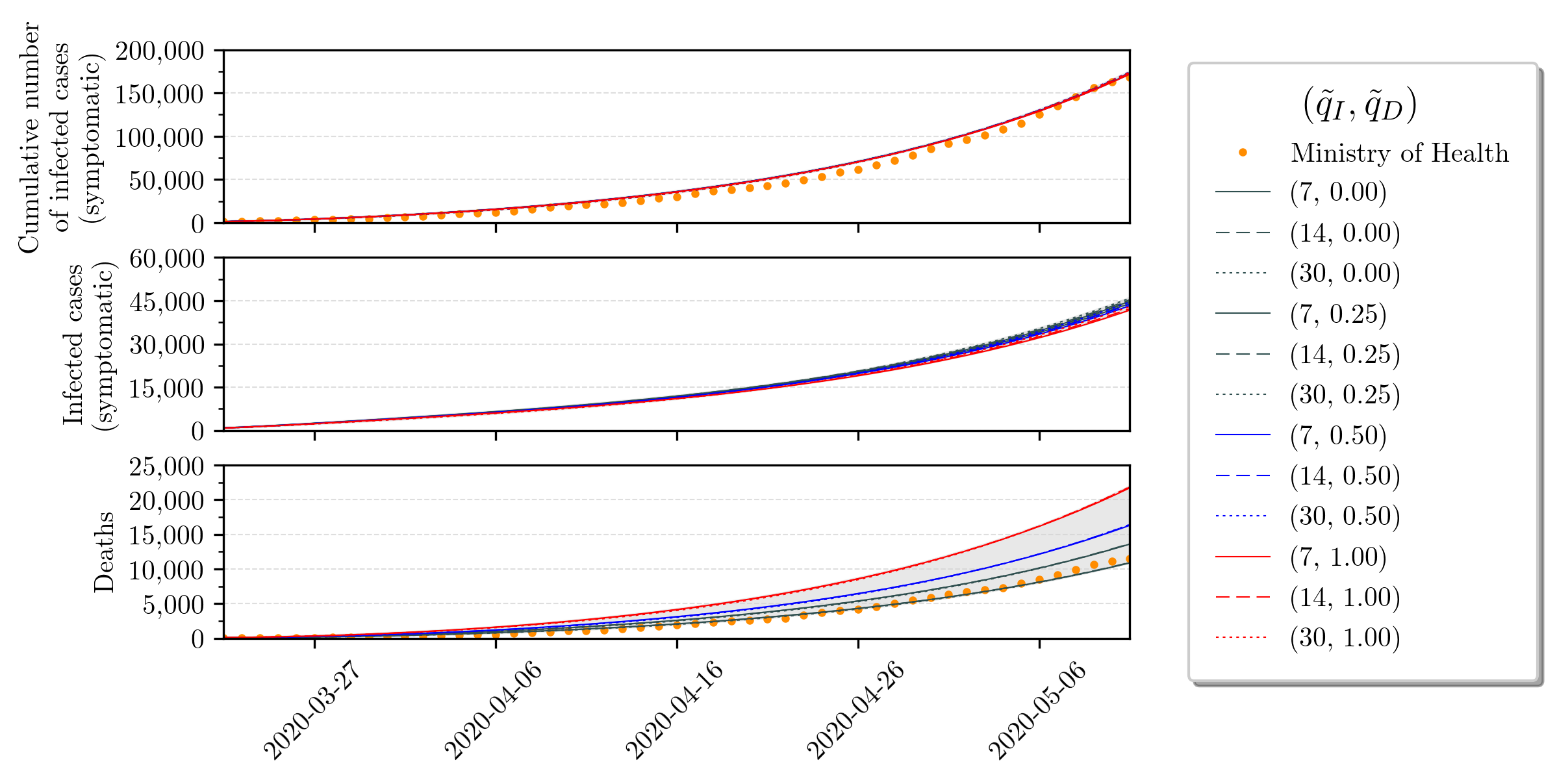}
\caption{Short-term simulation for the SIRASD$+\psi$ model using different values of uncertainty.}
\label{fig:simulation_sirasd_short}
\end{figure}

\begin{figure}
\centering
   \includegraphics[width=\linewidth]{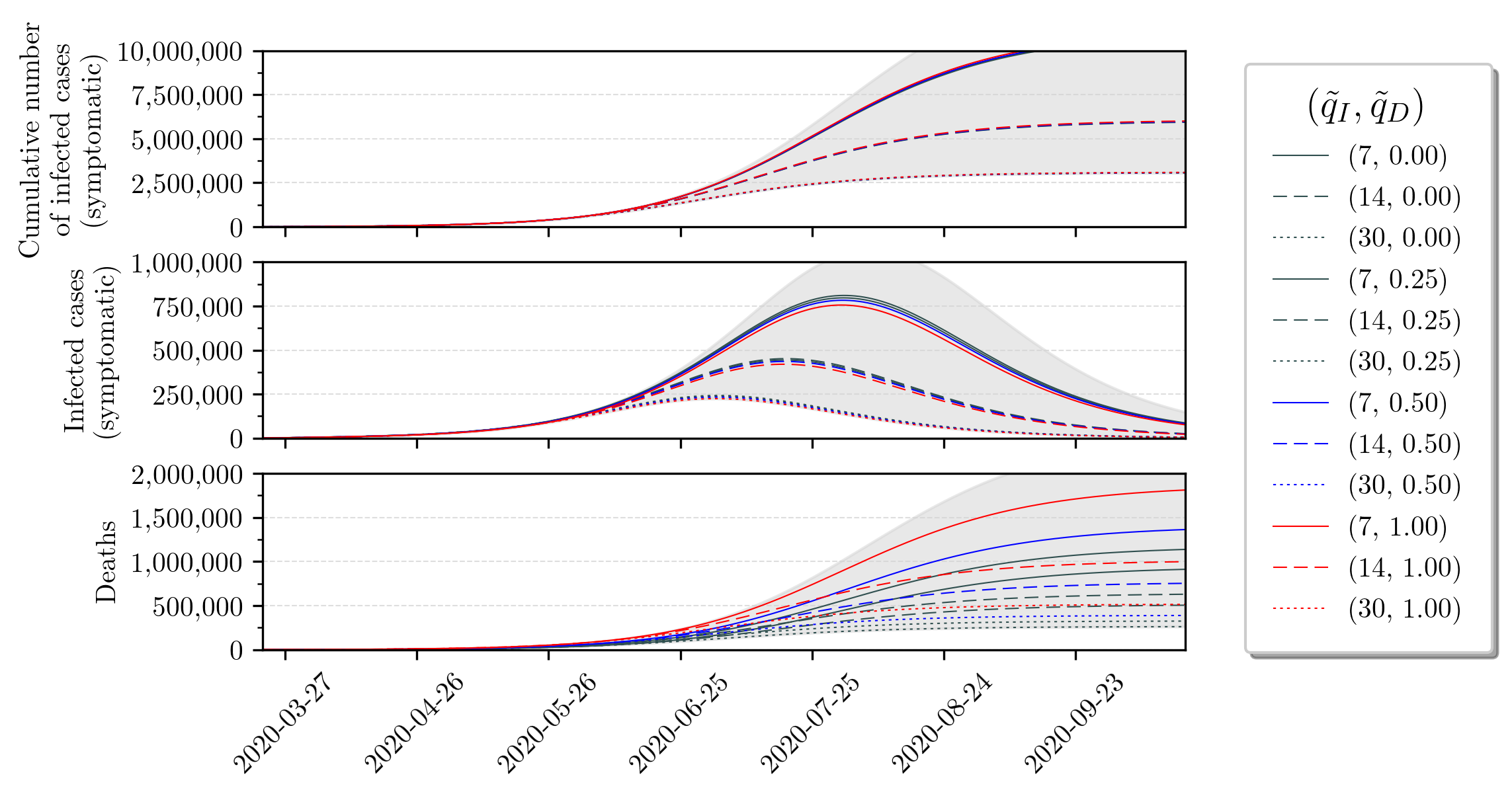}
\caption{Long-term simulation for the SIRASD$+\psi$ model using different values of uncertainty.}
\label{fig:simulation_sirasd_long}
\end{figure}

With respect to the uncertainty-embedded identification procedure in Section \ref{SecIdentif}, Figures \ref{fig:simulation_sirasd_short} and \ref{fig:simulation_sirasd_long} show, respectively,  short and long term predictions using the SIRASD$+\psi$ model. Table \ref{tab:simulation_info_sirasd} collects the essential information of these forecasts. Regarding the long-term forecast, we call attention to the infected cases. An increase in the infected uncertainty anticipates the peak in time, just as in the SIRD+$\psi$ model, but the symptomatic peak amplitude decreases. 

\begin{figure}
\begin{center}
\includegraphics[width=\linewidth]{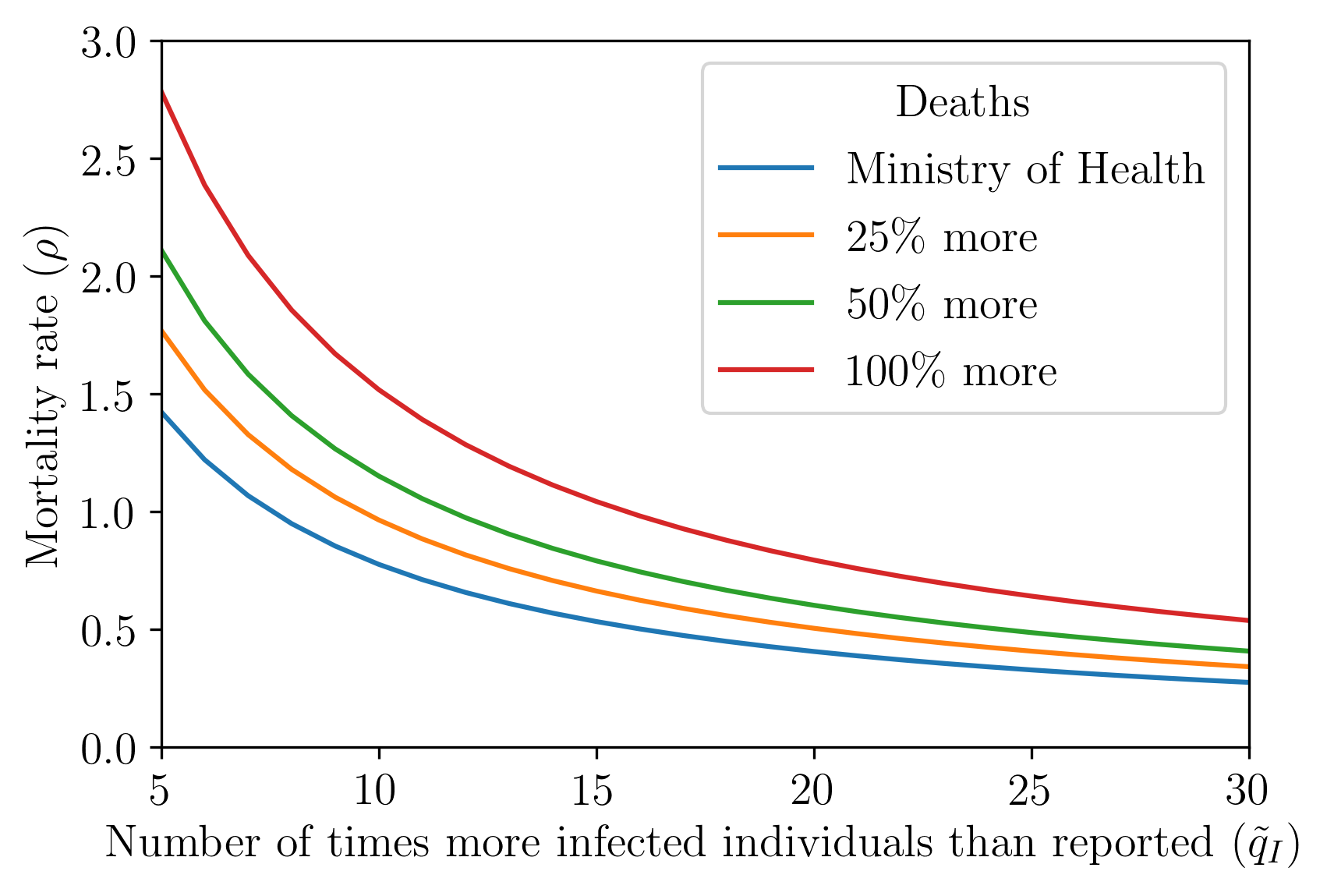}
\caption{\label{fig:mortality_sirasd}The number of times more infected than reported ($\tilde{q}_I$) vs. the observed mortality rate ($\rho$) for the SIRD$+\psi$ model. The blue, orange and green curves represent simulations using data from the Ministry of Health ($\tilde{q}_D = 0$), 50\% more deaths ($\tilde{q}_D = 0.5$) and 100\% more deaths ($\tilde{q}_D = 1$).}
\end{center}
\end{figure}

\begin{figure}
\centering
\includegraphics[width=\linewidth]{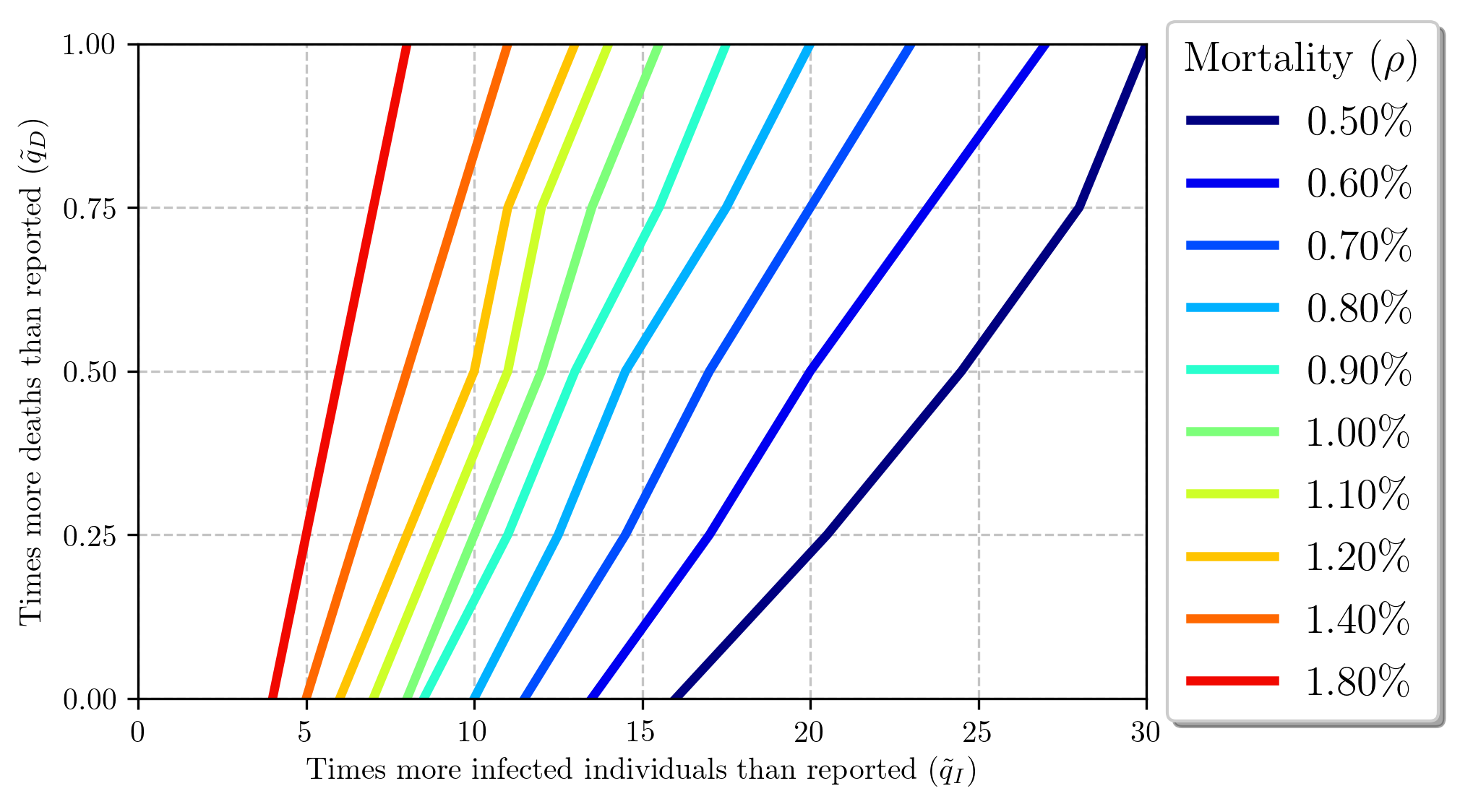}
\caption{Mortality rate paths using the number of times more infected than reported ($\tilde{q}_I$) versus the number of times more deaths than reported ($\tilde{q}_D$) for the SIRASD$+\psi$ model.}
\label{fig:sirasd_mortality_path}
\end{figure}

With respect to the simulated mortality rate, Figure \ref{fig:mortality_sirasd} shows the relationship between $\rho$ and the uncertainties $q_D$ and $q_I$; these values are detailed in Table \ref{tab:mortality_sirasd} of Appendix \ref{sec:appendix_observed_mortality_rates}. Results differ only slightly from those obtained with the SIRD+$\psi$ model and, thus, the same conclusions can be inferred. Furthermore, Figure \ref{fig:sirasd_mortality_path} gives an insight on the possible ``trajectories'' of $\rho$ with respect to these uncertainty margins, i.e. showing the static gain between $q_D$, $q_I$ and $\rho$, for different levels of mortality.

The identified epidemiological parameters for the SIRASD$+\psi$ model are presented in Table \ref{tab:parameters_sirasd}, in the Appendix. There are two interesting phenomena regarding $\beta_A$: (i) in all simulations we have $\beta_A > \beta_S$, meaning that asymptomatic individuals are more likely to transmit the disease since they probably do not know they are infected; and (ii) increasing the infected uncertainty causes a decrease in the parameter $\beta_A$, meaning that the probability to transmit the disease also decreases. As for $\gamma_A$, we notice that: (i) it is greater than $\gamma_S$ in all cases, meaning that asymptomatic individuals have a smaller infectious period ($1 / \gamma_A$); and (ii)  it decreases as the infected uncertainty increases, meaning that the average infectious period increases with the increases of asymptomatic individuals, i.e. it is more likely to find more asymptomatic individuals spreading the disease for a longer period.

\begin{figure}
\centering
\includegraphics[width=\linewidth]{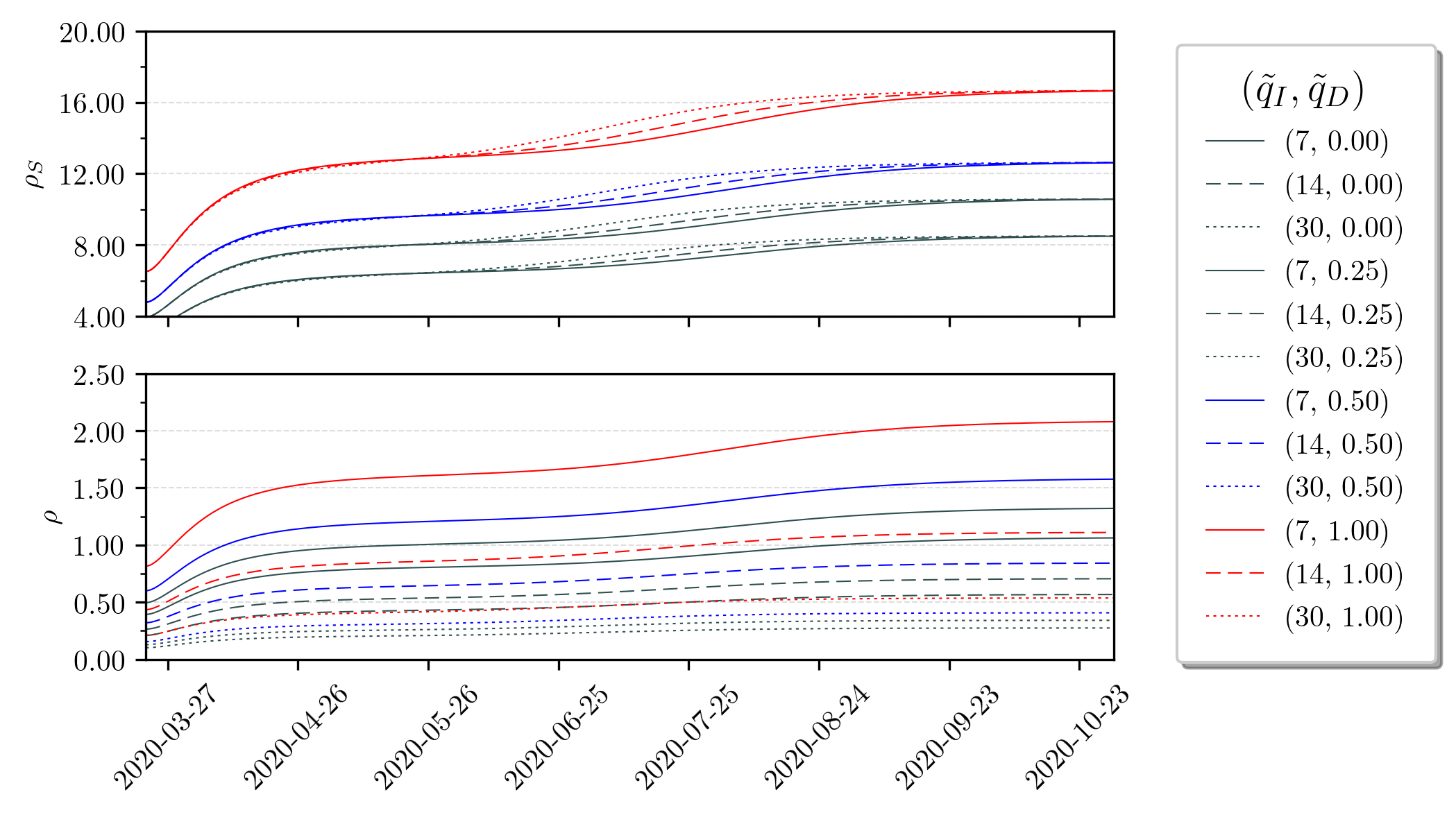}
\caption{Symptomatic and overall mortality rate over time for the SIRASD$+\psi$ model.}
\label{fig:SIRASD_rho_t}
\end{figure}

We must point out that analogous results to those achieved with the SIRD$+\psi$ model are found, as expected due to the equivalence between the nonlinear differential equations of these two models. We stress some key points:
\begin{itemize}
\item An increase on the sub-notification with respect to infected individuals directly implies in the reduction of the peak of symptomatic infections: essentially, the virus ``does not care'' if it causes symptoms or not on the infected individual (its only goal is to infect and replicate);
\item Assuming that the virus infects the same amount of people (despite the margins of the sub-reports), and as it infects (faster) more individuals without causing symptoms, we observe a smaller percentage peak of symptomatic infections for larger sub-notification.
\item The uncertainty margins upon the infected ($q_I$) and the death count ($q_D$) influence both infected and death curves, $I(t)$ and $D(t)$, respectively. The infected curve is directly and mostly influenced by the infected uncertainty, but the death uncertainty affects this curve more modestly through the total number of the population, $N(t)$, on the transmission channel (see Eq. (\ref{eq:SIRDead}) and Figure \ref{fig:simulation_sird_long}). Additionally, we emphasize that the deaths curve also depends on both uncertainty parameters, $q_D$ and $q_I$. The smallest death count forecast is found with an elevated under-reporting of asymptomatic infections and a small sub-report margins regarding deaths. This fact seems empirically reasonable, once an unaccounted increase on the number of deceased individuals should increase the number of deaths.

\item The instantaneous mortality rates, as of Eqs. (\ref{eq:rho_S(t)}) and (\ref{eq:rho_t}), are show in Figure \ref{fig:SIRASD_rho_t}. These instantaneous values indeed converge to those presented in Table \ref{tab:simulation_info_sirasd}. Anyhow, it is imperious to recall that the mortality rate, in practice, varies according to the amount of testing. Since the dynamics for symptomatic and asymptomatic infections evolve differently along time, the mortality rate also depends on the stage of the pandemic evolution. If the local epidemic scenario is an ending stage, the mortality rate tends to increase and stabilize. Note that if the margins of uncertainty are known (or roughly estimated), we can forecast quite accurately what will be the observed rates of mortality in the country.
\item The real mortality rate, measured with population samples, can be quite misleading, since it shows only an instantaneous snippet of the pandemic at a given moment (much like a "photograph"). Figure \ref{fig:SIRASD_rho_t} shows that the real mortality rate evolves in an asymptotic-like behaviour, converging to some steady-state value. Therefore, if one computes the mortality rate of a sample population in country with mass testing (unlike Brazil), and this country is not yet in an ending stage of its COVID-19 epidemic, one can observe values that are not the steady-state ones. In other words, there would still be symptomatic and asymptomatic to-be individuals which would alter the real mortality rate. And, as shows Figure \ref{fig:SIRASD_rho_t}, this real rate tends to be greater than in the beginning stages of the SARS-CoV-2 spread.
\end{itemize}

\begin{table}
\centering
\scriptsize
 \begin{tabular}{cccccc@{\extracolsep{3pt}}cc@{\extracolsep{3pt}}ccc}
\hline
~ & ~ & ~ & ~ & \multicolumn{2}{c}{Both} & \multicolumn{2}{c}{Symptomatic} & \multicolumn{2}{c}{Asymptomatic} & ~ \\ \cline{5-6} \cline{7-8} \cline{9-10}
$\tilde{q}_D$ & $\tilde{q}_I$ & $\rho_S$ & $\rho$ & Peak (\%) &Peak Forecast & Peak (\%) & Peak Forecast & Peak (\%) & Peak Forecast & Deaths \\ \hline
0 & 7 & 8.5408\% & 1.0676\% & 3.0667\% & 2020-07-31 & 0.3856\% & 2020-08-01 & 2.4267\% & 2020-07-31 & 939,784 \\
0 & 14 & 8.5335\% & 0.5689\% & 4.2421\% & 2020-07-18 & 0.2149\% & 2020-07-18 & 2.7655\% & 2020-07-18 & 512,176 \\
0 & 30 & 8.5208\% & 0.2749\% & 4.7137\% & 2020-07-03 & 0.1155\% & 2020-07-03 & 3.3167\% & 2020-07-03 & 262,242 \\
0.25 & 7 & 10.6162\% & 1.3270\% & 3.0667\% & 2020-07-31 & 0.3791\% & 2020-08-01 & 2.4505\% & 2020-07-31 & 1,172,135 \\
0.25 & 14 & 10.6073\% & 0.7072\% & 4.2421\% & 2020-07-18 & 0.2111\% & 2020-07-18 & 2.7858\% & 2020-07-18 & 638,423 \\
0.25 & 30 & 10.5882\% & 0.3416\% & 4.7137\% & 2020-07-03 & 0.1130\% & 2020-07-03 & 3.3205\% & 2020-07-03 & 325,994 \\
0.5 & 7 & 12.6686\% & 1.5836\% & 3.0761\% & 2020-07-31 & 0.3726\% & 2020-07-31 & 2.4751\% & 2020-07-31 & 1,403,689 \\
0.5 & 14 & 12.6580\% & 0.8439\% & 4.2421\% & 2020-07-18 & 0.2073\% & 2020-07-18 & 2.8063\% & 2020-07-18 & 764,006 \\
0.5 & 30 & 12.6358\% & 0.4076\% & 4.7117\% & 2020-07-03 & 0.1109\% & 2020-07-03 & 3.3414\% & 2020-07-03 & 390,023 \\
1 & 7 & 16.7050\% & 2.0881\% & 3.0761\% & 2020-07-31 & 0.3596\% & 2020-07-31 & 2.5249\% & 2020-07-31 & 1,864,283 \\
1 & 14 & 16.6919\% & 1.1128\% & 4.2494\% & 2020-07-18 & 0.1998\% & 2020-07-18 & 2.8485\% & 2020-07-18 & 1,013,367 \\
1 & 30 & 16.6639\% & 0.5375\% & 4.7117\% & 2020-07-03 & 0.1068\% & 2020-07-03 & 3.3830\% & 2020-07-03 & 516,959 \\
 \hline
\end{tabular}
 \caption{Observed mortality ($\rho_S$ and $\rho$), peak percentage and its respective date of occurrence and number of deaths after 360 days for the SIRD$+\psi$ model for different values of uncertainty.}
\label{tab:simulation_info_sirasd} 
\end{table}

\section{Discussion and Conclusions} 
\label{sec:discussion}

In this paper, we discussed the Brazilian COVID-19 pandemic scenario, the effects of under-reporting regarding the number of infections and deaths due to the SARS-CoV-2 virus in the country. Through two SIR-like adapted models, which include the population's response to measures of control of the pandemic (such as social isolation, the use of masks, etc.) and deaths due to the disease. Furthermore, we make a set of possible forecasts and evaluate how uncertainty meddles with the COVID-19 pandemic prediction curves. Specifically, we analysed how uncertainty affects the infection peak displacement in time and its amplitude, epidemiological parameters, the observed and real mortality rate evolution, and how infection and deaths uncertainty and the true mortality rate could be related.

We used two SIR-like models, SIRD$+\psi$ and SIRASD$+\psi$, which differ in two aspects. First, the SIRD$+\psi$ model considers that asymptomatic and symptomatic parameters are equal, that is, $\beta_S = \beta_A = \beta$ and $\gamma_S = \gamma_A = \gamma$ and that the mortality rate is applied in the whole infected population, while the SIRASD$+\psi$ model distinguishes symptomatic and asymptomatic parameters and applies the mortality rate to the symptomatic only. Second, the initial condition of the SIRD$+\psi$ model suggests that the infected class of individuals is actually split as $100q_I \%$ symptomatic and $100(1-q_I) \%$ asymptomatic, while the initial condition of the SIRASD$+\psi$ model considers that the proportion of one symptomatic per $\displaystyle \frac{(1 - q_I)}{q_I}$ asymptomatic individual (as gives Eq. \eqref{eq:d_measured_unknown_relationship}).

It is a fact (and heavily discussed by recent literature) that the available datasets disclosed by the Brazilian Ministry of Health have large sub-notification margins. Brazil currently conducts roughly $4000$ tests per million inhabitants, which is one of the lowest rates in the world \citep{volpatto2020spreading}. Likewise, the number of deaths may be underestimated for the same reason. According to \citep{covidsubnot}, only $8\,\%$ of infections are reported; the real number of infected individuals are possibly up to $30$ times bigger than what is being disclosed by the authorities \citep{scabini2020social,volpatto2020spreading}. 

Herein, we have tried to expose some essential insights regarding sub-notification and how to incorporate them to pandemic models; below, we summarize the main findings of this paper, enlightening the key points:
\begin{itemize}
    \item Since the spread of the SARS-CoV-2 virus is inherently complex and varies according to multiple factors (some which are possibly unmodelled and external), exact forecasts of the pandemic dynamics are not viable. Therefore, the correct procedure should be based on a recurrent model tuning (via identification), always taking into account the uncertainty margins. This measure would allow to present more coherent forecasts as time goes, since the uncertainty margins tend to decrease as the pandemic ceases (and as more testing is done). We remark that, in this paper, the uncertainty margins are assumed constant along the forecast horizons in order to tune/estimate the model parameters. In future works, the Authors plan in exploring the possible differences obtained when implying a dynamic behaviour for the uncertainty margins (asymptotic and decreasing).
    
    \item The simulation forecasts indicate that the amount of uncertainty influences directly on the date of the infection peak, on the number of deaths and on the mortality rate of the decease.
    
    Higher levels of infection uncertainty anticipates the infected peak in time. Considering no death uncertainty, the symptomatic peak at July 31, 2020 for $\tilde{q}_{I}=7$ is anticipated to July 3, 2020 for $\tilde{q}_{I}=30$ (see Table \ref{tab:simulation_info_sirasd}), almost a month difference. This is followed by the corresponding decrease (increase) in the peak amplitude of symptomatic (asymptomatic) individuals.
    
    Higher levels of death uncertainty cause increases in the observed and true mortality rates. Considering a fixed amount of infected uncertainty ($q_I$ or $\tilde{q}_I$ is constant), the number of deaths and the mortality rate are higher if the death uncertainty is higher (see Figures \ref{fig:simulation_sirasd_long} and \ref{fig:SIRASD_rho_t}). Also, considering a fixed amount of death uncertainty ($q_D$ or $\tilde{q}_D$ is constant), the number of deaths and the mortality rate are lower if the infected uncertainty is higher.
    
    Furthermore, there is a direct relationship between the uncertainty level and the observed mortality rates, which are in fact time-varying (Figure \ref{fig:SIRASD_rho_t}). The instant mortality rates are calculated using the number of deaths and the cumulative number of infected, while the true mortality rate is its asymptotic value. So depending in which stage of the epidemic we are at, estimations using population samples could vary, not only by the method itself, but also due to the natural evolution of the proportions of infected and deaths.

    \item Higher uncertainty levels of asymptomatic individuals causes a decrease in the epidemiological parameters $\beta_A$ and $\gamma_A$ (Table \ref{tab:parameters_sirasd}). There are two possible interpretations for the transmission parameter: (i) according to \citep{KEMPER1978707}, asymptomatic individuals presumably have more contacts, since they do not have any symptoms and, therefore, do not do a self-induced quarantine; and (ii) according to \citep{Robinson2013}, the transmission of the infection is more readily on symptomatic individuals due to physical signs of illness (coughing, sneezing, etc.), which outweighs this first factor. Since $\beta_A > \beta_S$, our results support the first hypothesis.
    
    That said, the transmission parameter $\beta_A$ is approximately 0.47 for $\tilde{q}_I = 5$ and 0.45 for $\tilde{q}_I = 30$, corresponding to 2.3\% to 6.8\% higher than the symptomatic transmission parameter\footnote{Considering $\beta_S = 0.44$, then $\beta_A = 0.47$ corresponds to $\displaystyle 0.03/0.44 \approx 6.8\%$ and $\beta_A = 0.45$ corresponds to $\displaystyle 0.01/0.44 \approx 2.3\%$.}, meaning that asymptomatic individuals are more likely to transmit the disease since the greater number of contacts over-weights the probability of transmitting the disease due to physical signs of illness.
    
    The duration of the infection $1 / \gamma_A$ is approximately $1 / 0.21 \approx 4.8$ days for $\tilde{q}_I = 5$ and $1 / 0.17 \approx 5.9$ days for $\tilde{q}_I = 30$, corresponding to 12\% to 28.4\% less than the average time to recovery of symptomatic individuals ($1 / \gamma_S = 1 / 0.15 \approx 6.7$ days), meaning that the infectious period of the asymptomatic is smaller than the symptomatic one. This is consistent with other pandemic, such as the H1N1 pandemic (see Table 1 from \cite{Robinson2013} for $\gamma_A$ and $\gamma_S$, and corresponding references).
    
    \item We find that the effect of under-reporting of the number of infected and deaths is related to the true mortality rate, which we call ``COVID-19 under-reporting tripod''. Considering that there are three variables with uncertainty (the under-reporting of infected individuals, under-reporting of deaths, and the true mortality rate), if two of them are known (or measured empirically), the other can be inferred assuming a constant amount of uncertainty. Alternatively, if one is known, it is possible to infer a path for the other two. This is shown in Figure \ref{fig:sirasd_mortality_path}.
    
    This approach allows to align the observed to the true mortality rate in order to find the population uncertainty. \citep{howDeadlyIsCoronavirus} mentions that many studies are estimating the true mortality rate to be in the range of 0.5-1\%. Assuming that this is the true  mortality rate, the number of times more infected than reported is about: 8-16 considering the deaths reported by the Ministry of Health; 10-20.5 considering 25\% more deaths than reported; 12-24.5 considering 50\% more deaths than reported; and 15.5-30 considering 100\% more deaths than reported. Alternatively, considering 9 times more infected than reported, the true mortality rate varies from 0.87\% to 1\%, resulting in 0 to 13\% more deaths than reported; if we consider 12 times more deaths than reported, the true mortality rate range is 0.67\%-1\%, and the death uncertainty range is 0-50\%. Other analysis can be inferred directly from Figure \ref{fig:sirasd_mortality_path}. These results are consistent with estimated and empirical findings presented in Table \ref{tab:estimated_underreporting}.

    \item We note that the level of uncertainty plays a significant role in shifting the Susceptible-Infected transmission curves of the contagion. Uncertainty not only influences the effective transmission rate per infectious contact, but also the recovery and mortality rates associated to the virus. It is possible to make a parallel between this paper and the recent study by \citep{Scarpino2020}, which investigates how macroscopic patterns and social reinforcement of interacting contagions through SIR-like models; it is shown how social reinforcement can directly meddle with the contagion transmission rate along the spread regime. The results point to very similar behaviours of the SIR curves as to the influence of the uncertainty levels. We note, nonetheless, that uncertainty also shifts and increases the peak of these curves, which does not happen in the demonstrations presented in \citep{Scarpino2020}.
\end{itemize}



We must stress, once again, that the used models have a series of limitations (such as unmodelled phenomena and disregarded transient behaviours), the available datasets are very imprecise, and also that the forecasting / model-based prediction problem has a lot of associated sensibility: it is composed of several coupled nonlinear differential equations, which heavily rely on initial conditions (and also contour factors due to the time-series behaviour implied via $\psi (t)$). Furthermore, the pandemic dynamics may vary abruptly if more intense health policies are adopted (or dropped) on the future. Therefore, we must recall that the reader that the results presented in this paper are qualitative. Our intentions in showing long-term predictions is not to provide perfect accurateness regarding the number of infection and deaths, but to show relevant phenomena regarding the levels of uncertainty. If more testing is performed, for example, the uncertainty levels tend to decrease and, thus, the forecast should also change. Also, if the elderly people were to be isolated from the virus in a more effective way, we would expect to see a decrease in death rate, even if the number of infected is increasing, since the mortality rate for this group of people is considerably higher than for people younger than 65 years, as shown by \citep{Perez-Saez2020.06.10.20127423}.

The Authors truly hope that the proposition herein formalised can serve to help determining adequate public health policies for Brazil.

\section*{Acknowledgment}
 D. O. C. and J. E. N. acknowledge the financial support of CNPQ under respective grants $302629/2019-0$ and $304032/2019-0$.
 
\subsection*{Notes}
The authors report no financial disclosure nor any potential conflict of interests.
 
\bibliographystyle{model5-names}
\bibliography{references}

\clearpage

\section{Appendix \label{sec:appendix}}

\subsection{Observed mortality rates \label{sec:appendix_observed_mortality_rates}}

\begin{table}[htp]
\centering
\small
 \begin{tabular}{ccccc}
\hline
\multirow{2}{*}{\text{Times more infected} ($\tilde{q}_I$)} & \multicolumn{4}{c}{Mortality rate ($\rho$)} \\ \cline{2-5}
 & $\tilde{q}_D = 0$ & $\tilde{q}_D = 0.25$ & $\tilde{q}_D = 0.5$ & $\tilde{q}_D = 1$ \\ \hline 
0 & 8.8569\% & 10.9324\% & 12.9481\% & 16.7947\% \\
1 & 4.6676\% & 5.8437\% & 7.0219\% & 9.3818\% \\
2 & 3.1651\% & 3.9819\% & 4.8088\% & 6.4925\% \\
3 & 2.3932\% & 3.0183\% & 3.6543\% & 4.9602\% \\
4 & 1.9233\% & 2.4292\% & 2.9457\% & 4.0113\% \\
5 & 1.6070\% & 2.0318\% & 2.4663\% & 3.3658\% \\
6 & 1.3798\% & 1.7458\% & 2.1207\% & 2.8988\% \\
7 & 1.2086\% & 1.5300\% & 1.8597\% & 2.5450\% \\
8 & 1.0750\% & 1.3614\% & 1.6555\% & 2.2677\% \\
9 & 0.9678\% & 1.2261\% & 1.4915\% & 2.0446\% \\
10 & 0.8799\% & 1.1151\% & 1.3568\% & 1.8612\% \\
11 & 0.8066\% & 1.0224\% & 1.2444\% & 1.7079\% \\
12 & 0.7444\% & 0.9438\% & 1.1490\% & 1.5777\% \\
13 & 0.6911\% & 0.8764\% & 1.0671\% & 1.4658\% \\
14 & 0.6447\% & 0.8177\% & 0.9958\% & 1.3683\% \\
15 & 0.6042\% & 0.7664\% & 0.9335\% & 1.2830\% \\
16 & 0.5684\% & 0.7211\% & 0.8784\% & 1.2077\% \\
17 & 0.5365\% & 0.6807\% & 0.8293\% & 1.1405\% \\
18 & 0.5080\% & 0.6446\% & 0.7854\% & 1.0803\% \\
19 & 0.4824\% & 0.6121\% & 0.7459\% & 1.0263\% \\
20 & 0.4591\% & 0.5826\% & 0.7100\% & 0.9771\% \\
21 & 0.4380\% & 0.5559\% & 0.6775\% & 0.9325\% \\
22 & 0.4187\% & 0.5315\% & 0.6478\% & 0.8917\% \\
23 & 0.4010\% & 0.5090\% & 0.6204\% & 0.8542\% \\
24 & 0.3847\% & 0.4884\% & 0.5953\% & 0.8198\% \\
25 & 0.3697\% & 0.4693\% & 0.5721\% & 0.7879\% \\
26 & 0.3557\% & 0.4517\% & 0.5507\% & 0.7585\% \\
27 & 0.3428\% & 0.4353\% & 0.5307\% & 0.7311\% \\
28 & 0.3307\% & 0.4200\% & 0.5121\% & 0.7055\% \\
29 & 0.3195\% & 0.4057\% & 0.4948\% & 0.6817\% \\
30 & 0.3090\% & 0.3924\% & 0.4785\% & 0.6594\% \\
\hline
\end{tabular}
 \caption{Observed mortality rates ($\rho$) for different simulations of the SIRD+$\psi$ model under infected ($\tilde{q}_I$) and death ($\tilde{q}_D$) uncertainty.}
\label{tab:mortality_sird} 
\end{table}

\begin{table}[htp]
\centering
\small
 \begin{tabular}{ccccc}
\hline
\multirow{2}{*}{\text{Times more infected} ($\tilde{q}_I$)} & \multicolumn{4}{c}{Mortality rate ($\rho$)} \\ \cline{2-5}
 & $\tilde{q}_D = 0$ & $\tilde{q}_D = 0.25$ & $\tilde{q}_D = 0.5$ & $\tilde{q}_D = 1$ \\ \hline 
0.1 & 7.9411\% & 9.8027\% & 11.6102\% & 15.2003\% \\
1 & 4.2743\% & 5.3126\% & 6.3393\% & 8.3585\% \\
2 & 2.8489\% & 3.5411\% & 4.2255\% & 5.5716\% \\
3 & 2.1364\% & 2.6554\% & 3.1687\% & 4.1782\% \\
4 & 1.7089\% & 2.1241\% & 2.5346\% & 3.3422\% \\
5 & 1.4239\% & 1.7698\% & 2.1119\% & 2.7848\% \\
6 & 1.2203\% & 1.5168\% & 1.8100\% & 2.3867\% \\
7 & 1.0676\% & 1.3270\% & 1.5836\% & 2.0881\% \\
8 & 0.9489\% & 1.1794\% & 1.4074\% & 1.8559\% \\
9 & 0.8539\% & 1.0614\% & 1.2665\% & 1.6701\% \\
10 & 0.7761\% & 0.9647\% & 1.1513\% & 1.5181\% \\
11 & 0.7114\% & 0.8842\% & 1.0552\% & 1.3915\% \\
12 & 0.6566\% & 0.8161\% & 0.9739\% & 1.2843\% \\
13 & 0.6096\% & 0.7578\% & 0.9043\% & 1.1924\% \\
14 & 0.5689\% & 0.7072\% & 0.8439\% & 1.1128\% \\
15 & 0.5333\% & 0.6629\% & 0.7910\% & 1.0431\% \\
16 & 0.5018\% & 0.6238\% & 0.7444\% & 0.9817\% \\
17 & 0.4739\% & 0.5891\% & 0.7030\% & 0.9270\% \\
18 & 0.4489\% & 0.5580\% & 0.6659\% & 0.8781\% \\
19 & 0.4264\% & 0.5300\% & 0.6325\% & 0.8341\% \\
20 & 0.4061\% & 0.5048\% & 0.6024\% & 0.7943\% \\
21 & 0.3876\% & 0.4818\% & 0.5749\% & 0.7581\% \\
22 & 0.3707\% & 0.4608\% & 0.5499\% & 0.7251\% \\
23 & 0.3552\% & 0.4415\% & 0.5269\% & 0.6948\% \\
24 & 0.3409\% & 0.4238\% & 0.5058\% & 0.6670\% \\
25 & 0.3278\% & 0.4075\% & 0.4862\% & 0.6412\% \\
26 & 0.3156\% & 0.3923\% & 0.4682\% & 0.6174\% \\
27 & 0.3043\% & 0.3783\% & 0.4514\% & 0.5953\% \\
28 & 0.2938\% & 0.3652\% & 0.4358\% & 0.5747\% \\
29 & 0.2840\% & 0.3530\% & 0.4212\% & 0.5555\% \\
30 & 0.2749\% & 0.3416\% & 0.4076\% & 0.5375\% \\
\hline
\end{tabular}
 \caption{Observed mortality rates ($\rho = p \, \rho_S$) for different simulations of the SIRASD+$\psi$ model under infected ($\tilde{q}_I$) and death ($\tilde{q}_D$) uncertainty.}
\label{tab:mortality_sirasd} 
\end{table}

\clearpage

\subsection{Simulation parameters \label{sec:appendix_simulation_parameters}}

\begin{table}[htb]
\centering
 \begin{tabular}{c}
\hline
\textbf{Number of deaths provided by the Ministry of Health ($\tilde{q}_D = 0$)} \\
\hline
\begin{tabular}{ccccc}
$\tilde{q}_I$ & $\beta$ & $\rho$ \\ \hline
0 & 0.441881 & 0.088569 \\
5 & 0.424779 & 0.016070 \\
10 & 0.423526 & 0.008799 \\
15 & 0.423275 & 0.006042 \\
20 & 0.423254 & 0.004591 \\
25 & 0.423415 & 0.003697 \\
30 & 0.423607 & 0.003090 \\
\end{tabular} \\
\hline
\textbf{25\% more deaths ($\tilde{q}_D = 0.25$)} \\
\hline
\begin{tabular}{ccccc}
$\tilde{q}_I$ & $\beta$ & $\rho$ \\ \hline
0 & 0.447062 & 0.109324 \\
5 & 0.425398 & 0.020318 \\
10 & 0.423706 & 0.011151 \\
15 & 0.423286 & 0.007664 \\
20 & 0.423174 & 0.005826 \\
25 & 0.423276 & 0.004693 \\
30 & 0.423427 & 0.003924 \\
\end{tabular} \\
\hline
\textbf{50\% more deaths ($\tilde{q}_D = 0.5$)} \\
\hline
\begin{tabular}{ccccc}
$\tilde{q}_I$ & $\beta$ & $\rho$ \\ \hline
0 & 0.452330 & 0.129481 \\
5 & 0.426041 & 0.024663 \\
10 & 0.423897 & 0.013568 \\
15 & 0.423304 & 0.009335 \\
20 & 0.423098 & 0.007100 \\
25 & 0.423139 & 0.005721 \\
30 & 0.423247 & 0.004785 \\
\end{tabular} \\
\hline
\textbf{100\% more deaths ($\tilde{q}_D = 1$)} \\
\hline
\begin{tabular}{ccccc}
$\tilde{q}_I$ & $\beta$ & $\rho$ \\ \hline
0 & 0.463086 & 0.167947 \\
5 & 0.427405 & 0.033658 \\
10 & 0.424317 & 0.018612 \\
15 & 0.423359 & 0.012830 \\
20 & 0.422956 & 0.009771 \\
25 & 0.422871 & 0.007879 \\
30 & 0.422889 & 0.006594 \\
\end{tabular} \\
\hline
\end{tabular}
 \caption{Estimated values of the epidemiological parameters for the SIRD$+\psi$ model. We used $\gamma = 0.150876$ provided by \citep{bastos2020modeling}, and $\alpha=0.186353$ and $\psi_{\infty}=0.494027$ (calculated with the data from the Ministry of Health, without any uncertainty) in all simulations.}
\label{tab:parameters_sird} 
\end{table}

\begin{table}
\centering
 \begin{tabular}{c}
\hline
\textbf{Number of deaths provided by the Ministry of Health ($\tilde{q}_D = 0$)} \\
\hline
\begin{tabular}{ccccccc}
$\tilde{q}_I$ & $\beta_A$ & $\beta_S$ & $\gamma_A$ & $\gamma_S$ & $\rho$ & $p$ \\ \hline
0.1 & 0.596899 & 0.441881 & 0.286665 & 0.150876 & 0.087352 & 0.909091 \\
1 & 0.495901 & 0.441881 & 0.206414 & 0.150876 & 0.085485 & 0.500000 \\
5 & 0.469934 & 0.441881 & 0.186263 & 0.150876 & 0.085431 & 0.166667 \\
10 & 0.464844 & 0.441881 & 0.182116 & 0.150876 & 0.085377 & 0.090909 \\
15 & 0.462002 & 0.441881 & 0.179598 & 0.150876 & 0.085324 & 0.062500 \\
20 & 0.459022 & 0.441881 & 0.177151 & 0.150876 & 0.085274 & 0.047619 \\
25 & 0.456489 & 0.441881 & 0.174976 & 0.150876 & 0.085225 & 0.038462 \\
30 & 0.453604 & 0.441881 & 0.172562 & 0.150876 & 0.085208 & 0.032258 \\
\end{tabular} \\
\hline
\textbf{25\% more deaths ($\tilde{q}_D = 0.5$)} \\
\hline
\begin{tabular}{ccccccccc}
$\tilde{q}_I$ & $\beta_A$ & $\beta_S$ & $\gamma_A$ & $\gamma_S$ & $\rho$ & $p$ \\ \hline
0.1 & 0.595876 & 0.447062 & 0.286665 & 0.150876 & 0.107830 & 0.909091 \\
1 & 0.488607 & 0.447062 & 0.200907 & 0.150876 & 0.106252 & 0.500000 \\
5 & 0.468361 & 0.447062 & 0.185085 & 0.150876 & 0.106190 & 0.166667 \\
10 & 0.463912 & 0.447062 & 0.181421 & 0.150876 & 0.106122 & 0.090909 \\
15 & 0.461257 & 0.447062 & 0.179043 & 0.150876 & 0.106060 & 0.062500 \\
20 & 0.458375 & 0.447062 & 0.176669 & 0.150876 & 0.105999 & 0.047619 \\
25 & 0.455899 & 0.447062 & 0.174537 & 0.150876 & 0.105939 & 0.038462 \\
30 & 0.453410 & 0.447062 & 0.172425 & 0.150876 & 0.105882 & 0.032258 \\
\end{tabular} \\
\hline
\textbf{50\% more deaths ($\tilde{q}_D = 0.5$)} \\
\hline
\begin{tabular}{ccccccccc}
$\tilde{q}_I$ & $\beta_A$ & $\beta_S$ & $\gamma_A$ & $\gamma_S$ & $\rho$ & $p$ \\ \hline
0.1 & 0.594797 & 0.452330 & 0.286665 & 0.150876 & 0.127712 & 0.909091 \\
1 & 0.481588 & 0.452330 & 0.195552 & 0.150876 & 0.126786 & 0.500000 \\
5 & 0.466812 & 0.452330 & 0.183913 & 0.150876 & 0.126715 & 0.166667 \\
10 & 0.462976 & 0.452330 & 0.180717 & 0.150876 & 0.126638 & 0.090909 \\
15 & 0.460515 & 0.452330 & 0.178486 & 0.150876 & 0.126566 & 0.062500 \\
20 & 0.457731 & 0.452330 & 0.176187 & 0.150876 & 0.126494 & 0.047619 \\
25 & 0.455307 & 0.452330 & 0.174095 & 0.150876 & 0.126425 & 0.038462 \\
30 & 0.452822 & 0.452330 & 0.171994 & 0.150876 & 0.126358 & 0.032258 \\
\end{tabular} \\
\hline
\textbf{100\% more deaths ($\tilde{q}_D = 1$)} \\
\hline
\begin{tabular}{ccccccccc}
$\tilde{q}_I$ & $\beta_A$ & $\beta_S$ & $\gamma_A$ & $\gamma_S$ & $\rho$ & $p$ \\ \hline
0.1 & 0.496673 & 0.463086 & 0.208421 & 0.150876 & 0.167203 & 0.909091 \\
1 & 0.468433 & 0.463086 & 0.185298 & 0.150876 & 0.167170 & 0.500000 \\
5 & 0.463762 & 0.463086 & 0.181556 & 0.150876 & 0.167088 & 0.166667 \\
10 & 0.461252 & 0.463086 & 0.179352 & 0.150876 & 0.166994 & 0.090909 \\
15 & 0.459041 & 0.463086 & 0.177363 & 0.150876 & 0.166901 & 0.062500 \\
20 & 0.456445 & 0.463086 & 0.175211 & 0.150876 & 0.166811 & 0.047619 \\
25 & 0.454125 & 0.463086 & 0.173201 & 0.150876 & 0.166724 & 0.038462 \\
30 & 0.451705 & 0.463086 & 0.171152 & 0.150876 & 0.166639 & 0.032258 \\
\end{tabular} \\
\hline
\end{tabular}
 \caption{Estimated values of the epidemiological parameters for the SIRASD$+\psi$ model. We used $\alpha=0.186353$ and $\psi_{\infty}=0.494027$ in all simulations.}
\label{tab:parameters_sirasd} 
\end{table}

\clearpage

\end{document}